\documentclass[aps, prl, twocolumn, notitlepage]{revtex4-1}
\usepackage{ulem}
\usepackage{xspace}
\usepackage{graphicx}
\usepackage{dcolumn}
\usepackage[usenames]{color}
\usepackage{slashed}
\usepackage[utf8]{inputenc}
\usepackage{amsfonts}
\usepackage{amsthm}
\usepackage{amsmath}
\usepackage{amssymb}
\usepackage{mathrsfs}
\usepackage{hyperref}
\usepackage{booktabs}
\usepackage{diagbox}
\usepackage{latexsym}
\usepackage{color}
\usepackage{bm}

\usepackage{float}
\makeatletter
\let\newfloat\newfloat@ltx
\makeatother
\usepackage{algorithm}
\usepackage{algorithmicx}
\usepackage{algpseudocode}

\newcommand{\printfnsymbol}[1]{%
  \textsuperscript{\@fnsymbol{#1}}%
}

\begin{document}
\title{Ground-state properties via machine learning quantum constraints}

\author{Pei-Lin Zheng\textsuperscript{\tiny $\#$}}
\affiliation{International Center for Quantum Materials, Peking University, Beijing, 100871, China}
\affiliation{School of Physics, Peking University, Beijing, 100871, China}
\author{Si-Jing Du\textsuperscript{\tiny $\#$}}
\affiliation{International Center for Quantum Materials, Peking University, Beijing, 100871, China}
\affiliation{School of Physics, Peking University, Beijing, 100871, China}
\author{Yi Zhang}
\email{frankzhangyi@gmail.com}
\thanks{\textsuperscript{\tiny $\#$} P.-L. Zheng and S.-J. Du are responsible for 1D fermion chains and spin chains respectively and contributed equally.}
\affiliation{International Center for Quantum Materials, Peking University, Beijing, 100871, China}
\affiliation{School of Physics, Peking University, Beijing, 100871, China}

\begin{abstract}
Ground-state properties are central to our understanding of quantum many-body systems. At first glance, it seems natural and essential to obtain the ground state before analyzing its properties; however, its exponentially large Hilbert space has made such studies costly, if not prohibitive, on sufficiently large system sizes. Here, we propose an alternative strategy based upon the expectation values of an ensemble of operators and the elusive yet vital quantum constraints between them, where the search for ground-state properties simply equates to classical constrained minimization. These quantum constraints are generally obtainable via sampling and then machine learning on a large number of systematically consistent quantum many-body states. We showcase our perspective on 1D fermion chains and spin chains for applicability, effectiveness, caveats, and unique advantages, especially for strongly correlated systems, thermodynamic-limit systems, property designs, etc.
\end{abstract}

\maketitle

\emph{Introduction}\textemdash The collective behaviors of quantum many-body systems are central to various cutting-edge fields in condensed matter physics and beyond. Despite the nominal simplicity of certain quantum Hamiltonians, e.g., the Hubbard model \cite{Scalettar1989,Scalapino1991,HubbardCompare2015}, the non-commuting quantum operators squander any advantageous basis, and the exponentially-large Hilbert space renders the solutions and characterizations of ground states costly, limiting the system size and geometry in numerical techniques, e.g., exact diagonalization and density matrix renormalization group (DMRG) \cite{SteveWhite1992, DMRG2005}. While quantum Monte Carlo methods introduce efficient samplings, they are limited to sign-problem-free cases \cite{Loh1990,QMCreview2001,Troyer2005}. Also, the ground state solution usually starts from scratch upon slight model modifications, making the systematic study of a complex phase diagram, not uncommon in condensed matter physics \cite{KivelsonIntertwine}, even more expensive.

Rather than the abstract quantum many-body ground state, we are usually interested in its properties, such as the ground-state energy and spontaneous-symmetry-breaking order parameters - (linear combinations of) expectation values of target observables. Considering that the minimum energy criteria also concern expectation values, one would be prompt to establish a study based solely on expectation values and cut out the ground state. However, the quantum operators follow nontrivial commutation relations and, as a result, enforce nontrivial quantum constraints upon their expectation values - a role played by the ground state as the mediator. Expectation values violating these quantum constraints do not have an underlying quantum state and may not reflect the true nature of the quantum many-body system. Therefore, such quantum constraints are complicated yet essential for proper expectation-value-based considerations. An example of such quantum constraints is the conformal bootstrap for conformal field theories \cite{CFTbootstrapRMP2019}, beyond which, however, the bootstrap reduces to mere bounds and no longer offers a controlled analysis of ground-state properties \cite{HanBootstrap, TQC2021}.

On the other hand, recent developments in machine learning \cite{MLbook, Lecun2015Deep} have revolutionized data analysis such as image recognition, spam and fraud detection, and autonomous driving \cite{Jordan2015Machine}. Artificial neural networks (ANNs) can grasp the key yet complex and hidden rules in big datasets and generalize accurately for future scenarios \cite{MLbook,Jordan2015Machine,Lecun2015Deep}. Recently, machine learning has witnessed many explorations at the quantum many-body physics frontier, including quantum state tomography \cite{Torlai2018,Torlai2019}, quantum phase recognition \cite{LeiWang2016,Melko20161,qlt2016,Kelvin2016,Simon2016,FrankMLZ2,Zhaihui2018,Lian2019}, neural network states \cite{Carleo2016,Deng2017}, experiment interpretations \cite{Melnikov2018,mlstm2019,Constas2020}, etc.

In this letter, we propose studying the ground-state properties of quantum many-body systems within a classical expectation-value framework with quantum constraints over an ensemble of important operators. Then, the ground-state properties amount to constrained minimization. We can generally encode such quantum constraints as ANNs via supervised machine learning on example quantum states. Without loss of generality, we showcase the unique advantages of our strategy on 1D fermion and spin-1/2 models: (1) Compared with the expensive procedure of solving quantum many-body states, evaluations of expectation values are efficient and easily parallelizable among multiple operators and quantum states. (2) Our main effort is to extract and apply the quantum constraints through a large, classical dataset of expectation values, where machine learning techniques excel compatibly and proficiently. (3) Given a sufficiently diverse and representative training set, the obtained quantum constraints work for all Hamiltonians with different parameters, where we iterate the classical constrained minimization with respect to the expectation values of different Hamiltonians. (4) We embed systematic properties such as system size, geometry, and dimensions into the sample quantum many-body states and are rarely limited by them. (5) The quantum constraints also exhibit the competition and symbiosis between the observables, offering recipes for engineering models for desired ground-state properties, even emergent phases.

\emph{Algorithm}\textemdash Our approach consists of steps as follows:
\begin{itemize}
    \item Start with a large and representative ensemble of quantum many-body states $\left\{\left|\Phi\right\rangle_{\alpha}\right\}$ systematically consistent with the potential ground state, namely, obeying the expected symmetries and the area law.
    \item For each $\left|\Phi\right\rangle_{\alpha}$, evaluate the expectation values of a set of operators $\{\hat O_j\}$ and contribute a physical data point $\langle {\bf \hat O}\rangle_{\alpha} = (\langle \hat O_1\rangle_{\alpha},\langle \hat O_2\rangle_{\alpha},\cdots)$ in the $\langle {\bf \hat O}\rangle$ space. Operators with lower orders and spatial extents receive priority due to larger relevance and compatibility with local Hamiltonians. For comparison, unphysical data is obtained by considering deviations from the physical data \cite{SuppGSPviaMLC}. 
   \item Via supervised machine learning on the training set $\{\langle {\bf \hat O}\rangle_{\alpha}\}$, train ANNs $f(\langle {\bf\hat O} \rangle)$ to distinguish physical (unphysical) values of $\langle {\bf \hat O}\rangle$ that is allowed (disallowed) by the quantum constraints.
   \item For the Hamiltonian $\hat{H}=\sum_j a_j \hat O_j$, search the constrained minimum of the energy $E=\sum_j a_j \langle \hat O_j \rangle$ with the quantum constraints $f(\langle {\bf\hat O} \rangle)$. The coordinates $\langle {\bf\hat O} \rangle_0$ of the resulting minimum offer the expectation values that characterize the ground state.
\end{itemize}

The first three steps yield the quantum constraints $f(\langle {\bf\hat O} \rangle)$ that mark the physical manifold in the classical $\langle {\bf \hat O}\rangle$ space. We expect a relatively smooth and continuous manifold, as the adiabatic theorem ensures that the quantum many-body ground states and the corresponding $\langle {\bf\hat O} \rangle$ evolve continuously in the absence of first-order phase transitions. Importantly, the area law and symmetries vastly reduce the pool of quantum many-body states from the original Hilbert space, and machine learning can summarize and generalize from a limited number of training samples \cite{MLbook, Lecun2015Deep}, making it feasible to extract the quantum constraints through a polynomial amount of sample states. Since evaluating expectation values is simple and efficient, the key is to obtain a diverse training set representative of the candidate parts of the Hilbert space, e.g., by teaming up multiple quantum many-body ansatzes.

Only the final step that applies the quantum constraints to the model Hamiltonians is repeated throughout a parameter space. Although the classical constrained optimizations are not guaranteed to be fully straightforward, compared with the exponential expense of brute-force quantum algorithms, the overall cost can be much less, especially given the available optimization algorithms and physical intuitions. For example, the solution $\langle {\bf\hat O} \rangle_0$ for one set of model parameters helps to initialize searches for its neighbors, as $\langle {\bf\hat O} \rangle_0$ changes continuously in the absence of transitions. For efficiency, we can start with models with exact solutions or controlled approximations \footnote{Such scenarios also provide valuable additional training samples.} and move progressively into other parts of the parameter space, tracking $\langle {\bf\hat O} \rangle_0$ successively in the process \cite{SuppGSPviaMLC}.

An important question is the choice of observables $\{\hat O_j\}$, for which we suggest two criteria: (1) Is the observable likely to appear in target Hamiltonians? (2) Does the observable represent a physical quantity we are interested in? These favor local, low-order operators, and the more, the better, though with added costs. Also, these criteria are soft: irrespective of chosen observables, quantum constraints address the physical realizability of their expectation values in a yes/no fashion; they do lose capacity without certain observables, e.g., tell corresponding degeneracy, but encounter no algorithmic breakdown.

\begin{figure}
\includegraphics[width=.9\linewidth]{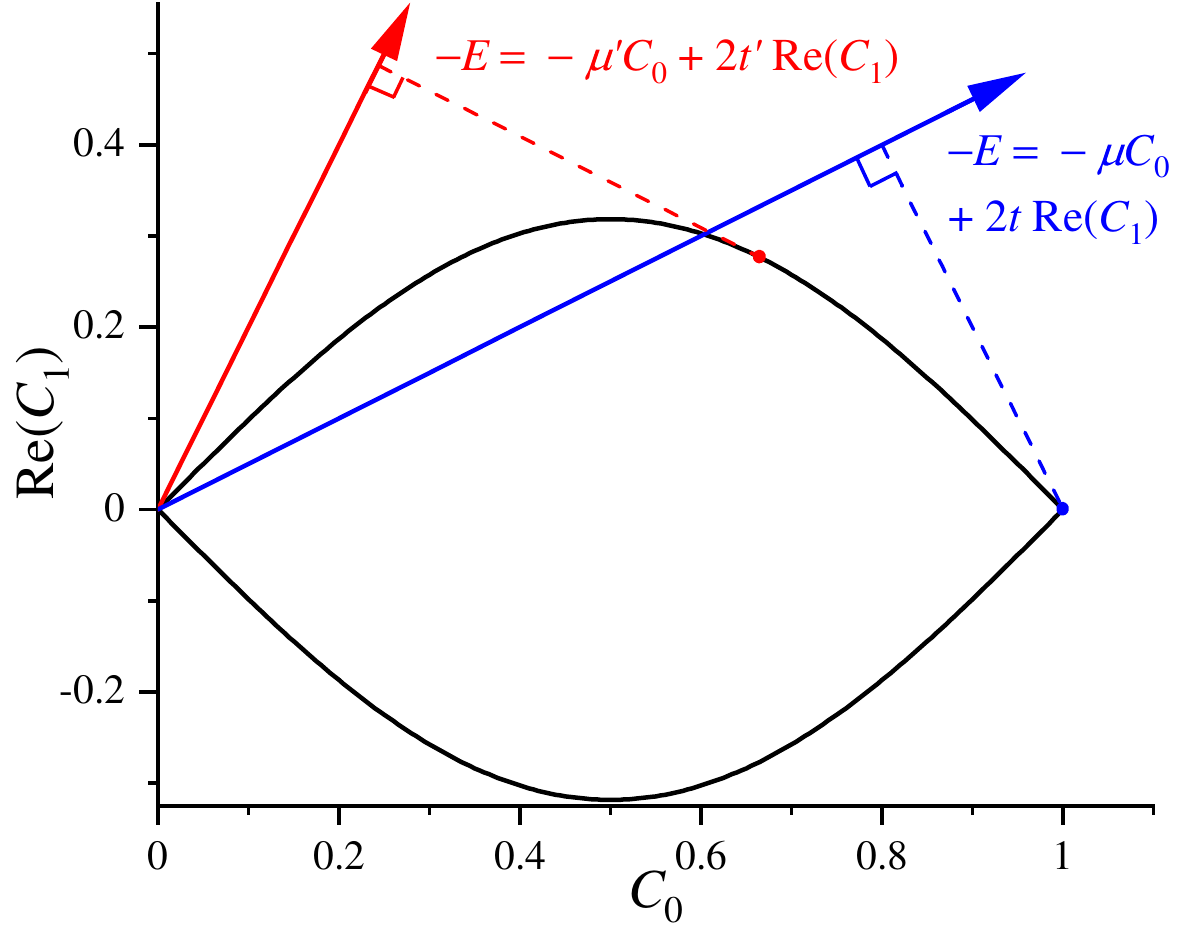}
\caption{The black contour shows the physical expectation values $(\mbox{Re}(C_1), C_0)$ consistent with the quantum constraint in Eq. \ref{eq:1dffC} for $\mbox{Im}(C_1) = 0$. Those expectation values not on the contour are unphysical, i.e., no quantum many-body state can realize them. Given different Hamiltonians, e.g., $t=1, \mu=-4$ (blue) and $t'=1, \mu'=-1$ (red), their energy expectation values correspond to projections along different directions. The coordinates of the physical point with the lowest energy characterize the ground-state properties.}\label{fig:fig1}
\end{figure}

\emph{A heuristic example}\textemdash First, let's consider a 1D Fermi sea between $k_L=k_0-k_F$ and $k_R=k_0+k_F$, where we have a simple analytical expression for the quantum constraints. Its expectation values of two-point correlators are:
\begin{eqnarray}
C_{0}&=&\left\langle c^\dagger_x c_x\right\rangle=k_{F}/\pi  \\ \nonumber
C_{r}&=&\left\langle c^\dagger_{x+r} c_x\right\rangle=\sin (k_{F} r)e^{ik_{0}r}/\pi r, r\ne 0,
\end{eqnarray}
irrespective of $x$ due to the translation symmetry. Expectation values of higher-order operators depend fully on $C_r$'s through Wick's theorem. In particular, the following quantum constraint holds between the most dominant real-valued $C_0$ and complex-valued $C_1$:
\begin{eqnarray}
\pm \pi \left|C_1\right| = \sin \left(\pi C_0\right), \label{eq:1dffC}
\end{eqnarray}
as illustrated in Fig. \ref{fig:fig1}, which holds as long as there is one Fermi sea and no spontaneous translation symmetry breaking. We note that the physical manifold (black contour) is smooth except for the two endpoints at $C_0 = 0, 1$, corresponding to Van Hove singularities. 

Now, let's consider a tight-binding Hamiltonian with nearest-neighbor hopping $t\in \mathbb{R}$ \footnote{The arguments can generalize to complex $t$ straightforwardly noting that the minimum energy requires $\arg{C_1}=-\arg{t}$.} and Fermi energy $\mu$:
\begin{eqnarray}
\hat{H}&=&\sum_{x}-t (c_{x+1}^{\dagger}c_{x} + c_{x}^{\dagger}c_{x+1})+\mu c_{x}^{\dagger}c_{x}\nonumber \\
E&=& \left(-2t \mbox{Re}(C_{1})+\mu C_{0}\right) \times N,
\label{eq:1dffE}
\end{eqnarray}
where $N$ is the system size, and we set $t=1$ as our unit of energy hereinafter. As the constraint in Eq. \ref{eq:1dffC} only concerns $|C_1|$, we set $\mbox{Im}(C_1)=0$ to allow maximal range for $\mbox{Re}(C_1)$. A schematic plot for the solutions of optimal values of $(\mbox{Re}(C_1), C_0)$ is in Fig. \ref{fig:fig1}. More rigorously, we define $\pi C_0 = y \in (0, \pi)$, and $\mbox{Re}(C_{1})/C_{0} = f(y)=\sin(y)/y \in(0,1)$ is a single-valued function following the quantum constraint. To minimize $E\propto [\mu-2t f(y)]\times y$, the subsequent solutions:
\begin{eqnarray}
2t f(y)&-&\mu + 2tf'(y)y=0 \Rightarrow 2t\cos(y)=\mu \nonumber \\
C_0 &=& y/\pi = \arccos(\mu/2t)/\pi \\
\mbox{Re}(C_1) &=& yf(y) /\pi = \sin(y)/\pi= \mbox{sgn}(t)\sqrt{1- \mu^2 / 4t^2} /\pi, \nonumber
\end{eqnarray}
which are consistent with the exact results obtained in the momentum space $H=\sum_k [-2t\cos(k)+\mu]c^\dagger_k c_k$ for generic values of $t$ and $\mu$.

For ground-state properties of Hamiltonians with upto $n^{th}$-nearest-neighbor hopping, we need to employ quantum constraint $C_0 = f(C_1/C_0, C_2/C_0, \cdots)$ on the expectation values $C_i$, $i = 0,1,2,\cdots,n_{fs}$, which can be represented by ANNs and trained via supervised machine learning on quantum states with multiple Fermi seas \cite{SuppGSPviaMLC}. For general quantum many-body systems, we may not formulate the quantum constraints as a function between the expectation values. It is more convenient to establish a `penalty' function $f(\langle {\bf \hat O} \rangle)$ that measures the extent of $\langle {\bf \hat O} \rangle$'s violations to the quantum constraints \cite{SuppGSPviaMLC}, which is also advantageous for allowing more freedom in choices of $\bf \hat O$. We will examine such formalism next.

\emph{Benchmark examples}\textemdash We consider 1D fermion insulators with a bipartite unit cell, whose Bloch states take a general form $u(k)=\left(\cos(\theta_k/2), \sin(\theta_k/2) \exp(i\varphi_k)\right)^T$ with the first (second) component denoting the A (B) sublattice. The expectation values of two-point correlators are:
\begin{eqnarray}
C^{AA (BB)}_{0}&=&0.5\pm g_0/2, \nonumber \\
C^{AA (BB)}_{r}&=& \pm g_r/2, r\in \mathbb{Z}^{+}, \nonumber\\
C^{AB}_{r'}&=&\tilde g_{r'}/2, r'\in \mathbb{Z}+1/2, \label{eq:Ctog}
\end{eqnarray}
the rest obtainable via complex conjugation. $g_r =  \int^{2\pi}_0 \frac{dk}{2\pi} e^{ikr} \cos(\theta_k)$, $\tilde g_{r'} =  \int^{2\pi}_0 \frac{dk}{2\pi} e^{i(kr'+\varphi_{k})} \sin(\theta_{k})$, over which we establish the following quantum constraints:
\begin{equation}
\sum_r g_r \cdot g^*_{r+s} + \sum_{r'} \tilde g_{r'} \cdot \tilde g^*_{r'+s} = \delta_s.
\label{eq:1dmfC}
\end{equation}
$g_r$ and $\tilde g_{r'}$, related to correlations in insulators, are fast decaying functions of $r$ and $r'$, allowing us to truncate at a finite distance $\Lambda=20$ unless noted otherwise. We can thus define a positive-definite penalty function:
\begin{eqnarray}
f(g_r, \tilde g_{r'}) = \sum_{s=0}^{\Lambda/2} \left[ \sum_r g_r \cdot g^*_{r+s} + \sum_{r'} \tilde g_{r'} \cdot \tilde g^*_{r'+s} - \delta_s   \right]^2,
\label{eq:distophys}
\end{eqnarray}
which yields $\sim 0$ if and only if $\{g_r,\tilde g_{r'}\}$ are consistent with the quantum constraints. We note that the derivation of an expression as Eq. \ref{eq:distophys} is unavailable in generic quantum scenarios. Here for non-interacting fermions, it offers benchmarks to our strategy via machine learning quantum constraints in the following paragraphs.

Starting from random $u(k)$, we obtain $1.92\times 10^6$ samples of $\{g_r, \tilde{g}_{r'}\}$ consistent with the quantum constraints and no penalty. We also include in the dataset $7.68\times 10^6$ contrasting samples with small random deviations to $\{g_r, \tilde{g}_{r'}\}$ and corresponding penalties \cite{SuppGSPviaMLC}. Besides, we utilize the gauge equivalence to reduce the degrees of freedom \cite{SuppGSPviaMLC}. Then, we apply supervised machine learning \cite{MLbook, Lecun2015Deep} to train ANNs on the quantum constraints of $\{g_r,\tilde g_{r'}\}$ in the neighborhood of small or no violations \cite{SuppGSPviaMLC}. In practice, we use the average output of multiple independent ANNs $f^{*}(g_r, \tilde g_{r'})$ as the approximate penalty and their max output as an acceptance threshold to avoid unphysical regions.

To test out these quantum constraints, we study the mean-field solutions of a 1D interacting fermion Hamiltonian at half-filling:
\begin{eqnarray}
\hat{H}=\sum_{x}-t (c_{x+1}^{\dagger}c_{x}+c_{x}^{\dagger}c_{x+1}) + V c^\dagger_{x+1}c_{x+1}c^\dagger_x c_x. \label{eq:1Dins}
\end{eqnarray}
The underlying assumptions of $f(g_r, \tilde g_{r'})$ and $f^{*}(g_r, \tilde g_{r'})$ are that the ground state takes a non-interacting fermion framework, hence the Hartree-Fock approximation, and an emergent bipartite order parameter may spontaneously break the translation symmetry. Likewise, while our strategy straightforwardly applies to any quantum many-body ansatz, e.g., matrix product states \cite{MPSreview2008, Vidal2007, Vidal2008}, neural network states \cite{Carleo2016, Deng2017}, quantum Monte Carlo methods, ab initio wave functions, even multiple ansatzes at the same time, the resulting quantum constraints will inherit the underlying assumptions and skip lower-energy scenarios beyond such assumptions, if any.

Under these circumstances, the energy expectation value is:
\begin{eqnarray}
E&=&\left\langle \hat{H}\right\rangle=\left[ -t \left(\mbox{Re} (\tilde g_{1/2}) + \mbox{Re} (\tilde g_{-1/2}) \right) \right.  \nonumber \\
&+& \left. 0.25V\left(2-2g_0^2-|g_1|^2-|g_{-1}|^2 \right) \right]\times N/2. \label{eq:1dmfE}
\end{eqnarray}
We look for the constrained minimum $\{g_r, \tilde{g}_{r'}\}$ by minimizing either $L=\bar E +\eta f$ or $L=\bar E +\eta f^{*}$, where $\bar E = E/(N/2)$. $\eta$ controls the weight of the quantum constraints, and the optimized results approach the physical limit asymptotically when $\eta\rightarrow\infty$. In practice, we should balance $\eta$ between too large to allow an efficient search acceptance rate and too small to prevent the search from exiting regions represented by the samples. The extrapolation of $\eta$ may offer a more systematic analysis, and an example is shown in Supplemental Material \cite{SuppGSPviaMLC}.

\begin{figure}
\includegraphics[width=1.0\linewidth]{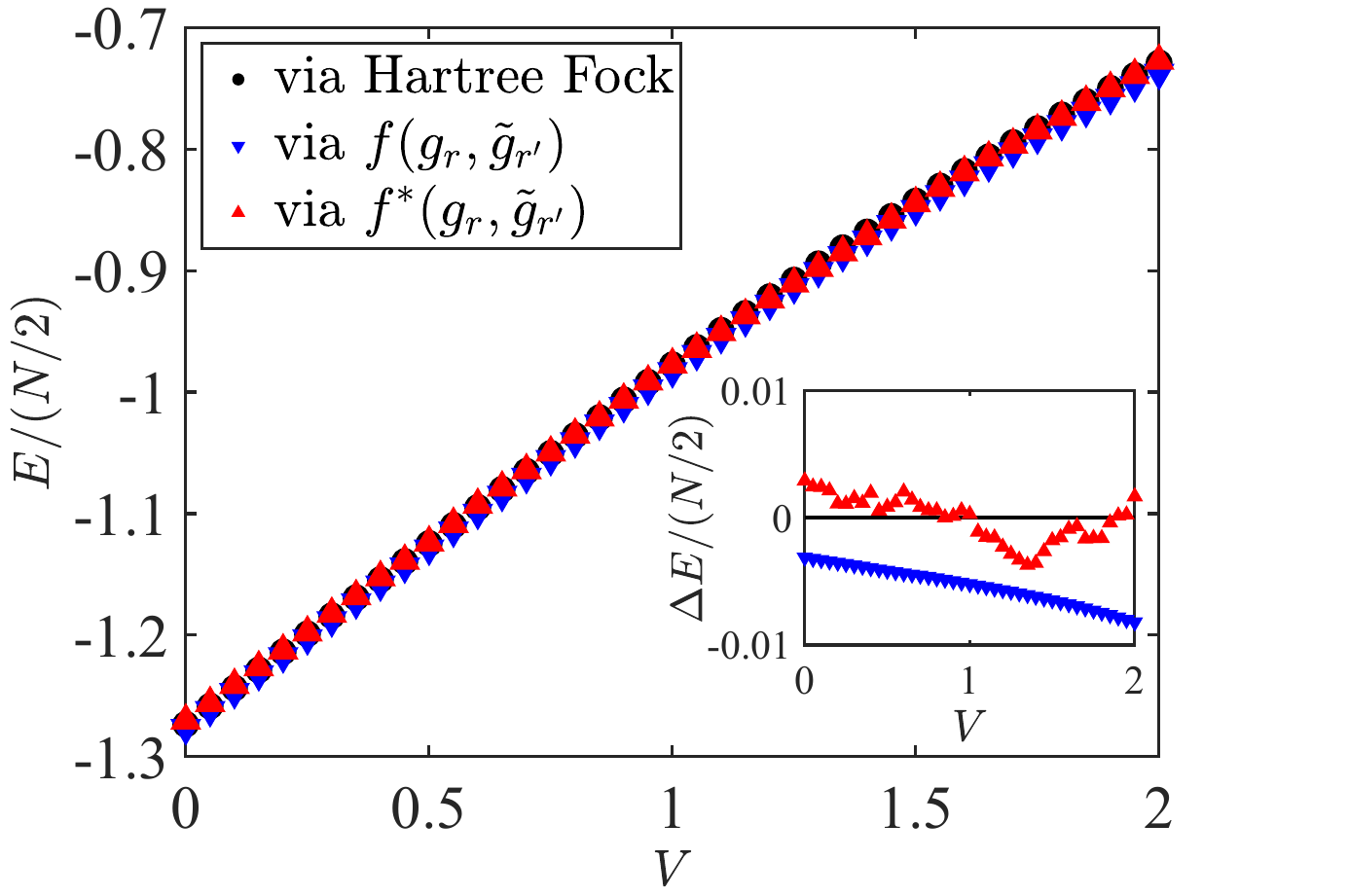}
\includegraphics[width=1.0\linewidth]{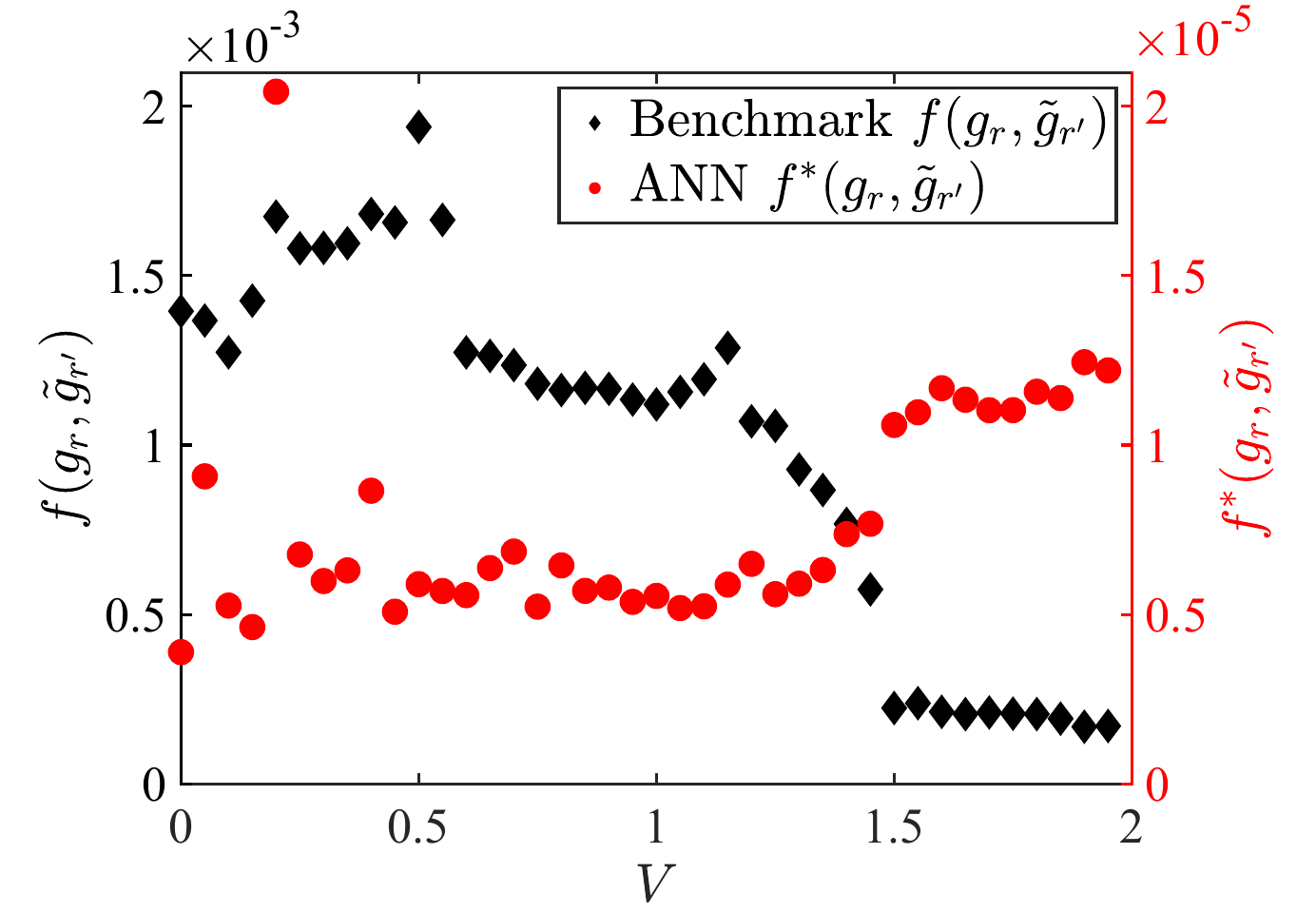}
 \caption{Top: The ground-state energies and (inset) their relative difference of the quantum many-body system in Eq. \ref{eq:1Dins} from the Hartree Fock approximation as well as the constrained minimization of $L=\bar E+\eta f$ and $L=\bar E+\eta f^*$, respectively. $\eta=1000$. Note that the energies obtained with our strategy may end up slightly below the theoretical values since we have slacked the quantum constraints for improved efficiency. Bottom: The ANN outputs $f^{*}(g_r, \tilde g_{r'})$ for the constrained minimum as well as the benchmark $f(g_r, \tilde g_{r'})$ for the same $\{g_r, \tilde{g}_{r'}\}$ show consistency with the quantum constraints (very small penalty values) throughout the $V$ range as we gradually lower from $V=2$.}\label{fig:CDWvsV}
\end{figure}

Also, we use the expectation values of $\{g_r, \tilde{g}_{r'}\}$ at $V$ to initialize the search at $V+\delta V$, and so on so forth. In practice, we start from $V=2$ with an interval of $\delta V=-0.01$ \footnote{Although the model at $V=0$ also offers a good starting point, the truncation in $r$ and $r'$ for a metal state may cause issues.}. The benchmark results are summarized in Fig. \ref{fig:CDWvsV} and Ref. \cite{SuppGSPviaMLC}, and their consistency indicates that given sufficient dataset and training, machine learning can offer a trustworthy path toward quantum constraints. It is worth noting that such soft quantum constraints offer a distinctive and complementary perspective to conventional variational approaches: while the latter bounds the ground states from above, given a search space generally smaller than necessary, our method may approach the ground state from below, where near-physical regions join our consideration yielding a search space larger than permitted.

\begin{figure}
	\includegraphics[width=0.95\linewidth]{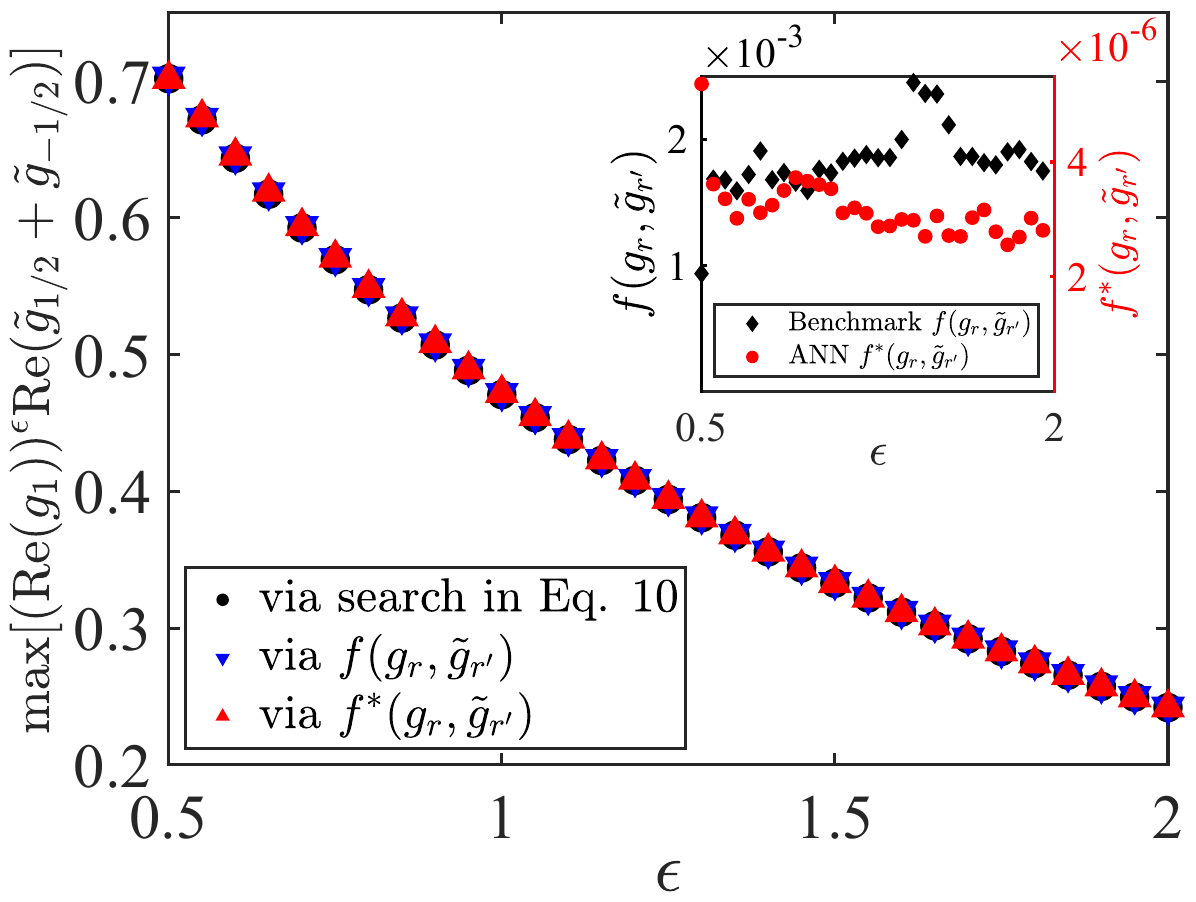}
	\caption{The maximum of $(\mbox{Re}(g_1))^\epsilon \mbox{Re}(\tilde g_{1/2} + \tilde g_{-1/2})$ for 1D fermion insulators on bipartite lattices shows good consistency between constrained optimization via the quantum constraints and searches among the variational Hamiltonian $H_{var}$ in Eq. \ref{eq:searchHam}. Individual expectation values for realizing the maximum, e.g., $g_1$ and $\tilde g_{1/2}$, also check out for each $\epsilon$. For the quantum-constraints approaches, we gradually increase $\epsilon$ from $\epsilon=0.5$ while keeping track of $\{g_r, \tilde{g}_{r'}\}$. Inset: the ANN $f^{*}(g_r, \tilde g_{r'})$ and the benchmark $f(g_r, \tilde g_{r'})$ suggest the obtained $\{g_r, \tilde{g}_{r'}\}$ indeed obey the quantum constraints. $\eta=1000$.}\label{fig:design}
\end{figure}

The quantum-constraint perspective also allows us to design quantum many-body systems like never before. Say we wish to apply certain criteria to expectation values: sometimes it is as simple as the inclusion of the corresponding observables into the Hamiltonian, yet sometimes the criteria do not possess simple interpretations or require a nontrivial origin such as spontaneous symmetry breaking. For instance, to maximize $(\mbox{Re}(g_1))^\epsilon \mbox{Re}(\tilde g_{1/2} + \tilde g_{-1/2})$, $\epsilon\in[0.5,2]$ for 1D fermion insulators on bipartite lattices, we commonly need to consider a variational Hamiltonian, such as:
\begin{equation}
    H_{var}=\sum_{x}-t (c_{x+1}^{\dagger}c_{x}+\text{h.c.}) + (-1)^x \Delta (c_{x+2}^{\dagger}c_{x}+\text{h.c.}), \label{eq:searchHam}
\end{equation}
which balances operators favoring $\mbox{Re}(\tilde g_{1/2} + \tilde g_{-1/2})$ and $\mbox{Re}(g_1)$, respectively, and $\Delta$ is a variational parameter for optimization. More generally, larger variational spaces with additional operators are preferred for thorough searches, and the ground-state solutions may bring additional complications. On the other hand, with the quantum constraints, we can circumvent such difficulties and resort to a constrained maximization. We compare our results in Fig. \ref{fig:design}. Further, we can establish the underlying Hamiltonians and quantum states via the strategy in Ref. \cite{Jiabao2022}.

\emph{Strongly correlated scenarios}\textemdash Generality is another essential merit of our strategy, which applies straightforwardly to strongly correlated systems and outshines the conventional methods hanging on the mind-boggling quantum many-body ground states themselves. Here, we illustrate the quantum constraints of 1D interacting spin-$1/2$ chains in the thermodynamic limit. Since the ground states of local Hamiltonians obey the Area Law, we can use tensor network states \cite{SteveWhite1992, MPS1992, DMRG2005,MPS-DMRG}, especially infinite matrix product states \cite{MPSreview2008, Vidal2007, Vidal2008} for infinite system sizes, as our representation of quantum many-body state samples for machine learning quantum constraints. We also emphasize that it is straightforward to generalize and beneficial to include other quantum may-body ansatzes, such as projected states via variational Monte Carlo methods; see Supplemental Materials for examples and results \cite{SuppGSPviaMLC}.

First, we sample quantum many-body states with random, translation symmetric matrices of dimension $\chi=8$ \footnote{We note that random matrix product states have emergent statistics \cite{RMPS1, RMPS2, RMPS3} against sampling generality at large dimension $\chi$, hence our small $\chi$ choice.}. Next, we evaluate the expectation values $\langle \hat{\bf O}\rangle$ of a series of low-order spin operators $S^\lambda_r$ and $S^\lambda_r S^{\lambda'}_{r+l}$, $\lambda, \lambda'=x,y,z$ upon a section of length $l_{max}=6$. Other than $6.75\times 10^4$ of these quantum-constraints-abiding samples, we also include $2.025\times 10^5$ contrasting samples with small deviations and corresponding penalties \cite{SuppGSPviaMLC}. Then, we perform supervised machine learning on the dataset and train ANNs $f^{*}(\langle \hat{\bf O}\rangle)$ to recognize how well a target $\langle \hat{\bf O}\rangle$ aligns with the quantum constraints. We note that the trained ANNs, as well as the previous ANNs $f^{*}(g_r,\tilde{g}_{r'})$ and benchmark $f(g_r,\tilde{g}_{r'})$ for 1D fermion insulators, penalize expectation values' departure from and thereby enforcing quantum constraints as intended \cite{SuppGSPviaMLC}.

\begin{figure}
	\includegraphics[width=0.95\linewidth]{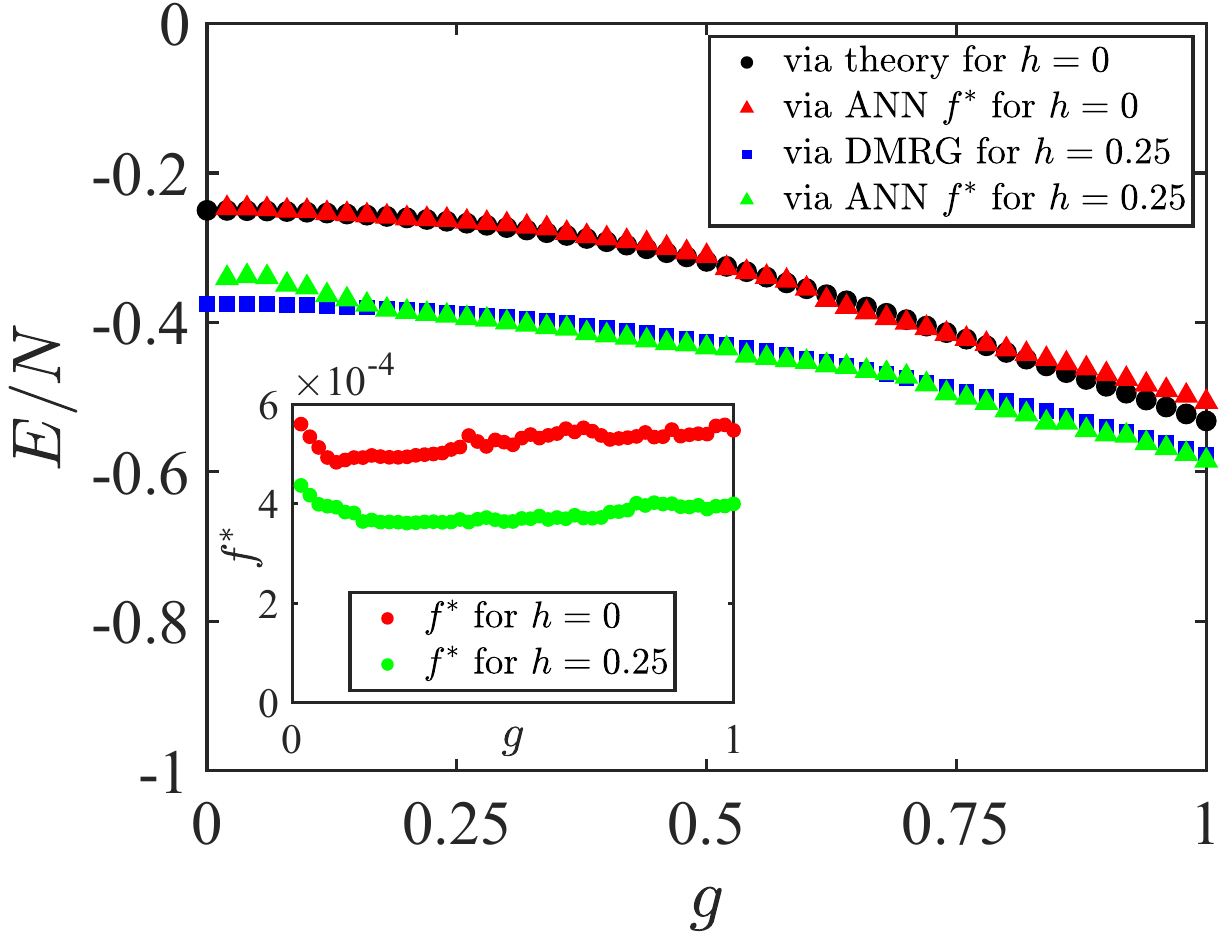}
	\caption{The ground-state energies of the transverse field Ising model ($h=0$) and longitudinal-transverse field Ising model ($h=0.25$) in Eq. \ref{eq:1dtfim} obtained by theory solutions \cite{LIEB1961}, DMRG, and the constrained minimization using ANN quantum constraints $f^{*}(\langle \hat{\bf O}\rangle)$ exhibit satisfactory consistency. For the latter, we start from $g=0$ and gradually increase $g$. $\eta=1000/3$ for $h=0$ and $\eta=400$ for $h=2.5$, respectively. Inset: small $f^{*}(\langle \hat{\bf O}\rangle)$ suggests that the obtained results well satisfies the quantum constraints.} \label{fig:GSE_TFI}
\end{figure}

Subsequently, we use the quantum constraints for ground-state properties of quantum spin Hamiltonians. For instance, we apply our strategy with $f^{*}(\langle \hat{\bf O}\rangle)$ to the 1D transverse field ($h=0$) and the non-integrable longitudinal-transverse field Ising models ($h\neq 0$):
\begin{eqnarray}
H=\sum_j -J S^z_{j} S^z_{j+1} - g S^x_j - h S^z_j,
\label{eq:1dtfim}
\end{eqnarray}
where wet set $J=1$ as the unit of energy. The results on the energy expectation value per site $E/N=\langle H \rangle / N = -J \langle S^z_{0} S^z_{1}\rangle - g \langle S^x_0 \rangle - h \langle S^z_0 \rangle $ and beyond in the $N \rightarrow \infty$ thermodynamic limit are summarized in Fig. \ref{fig:GSE_TFI} and Supplemental Material \cite{SuppGSPviaMLC}. Such quantum constraints are also directly applicable to the spin-$1/2$ XXZ chains \cite{SuppGSPviaMLC}. In summary, we obtain quantitative results on ground-state properties such as energies and short-range correlators yet qualitative trends only on longer-range correlators, making pinpointing phase transitions relatively tricky. An additional or different set of observables addressing critical behaviors may be helpful.

\emph{Discussions}\textemdash We propose to analyze ground-state properties via machine learning quantum constraints on expectation values and complement conventional ground-state-based approaches. Other than the aforementioned advantages, we have yet to establish a controlled quantitative analysis of algorithmic uncertainties, especially for relatively soft degrees of freedom, e.g., the order parameter of a spontaneous symmetry-breaking phase. Qualitative tendencies are observable, and extrapolation of $\eta$, the weight of quantum constraints, offers a partial solution \cite{SuppGSPviaMLC}. Also, tensor network states in 2D and beyond may become costly, and other quantum many-body ansatzes may help complement the training data and simultaneously reduce biases originating from respective ansatz. Finally, while systematic presumptions such as the area law and symmetries help narrow the questions and facilitate the calculations, such physics intuitions should sometimes be taken with a grain of salt.

\emph{Acknowledgement:} We thank Junren Shi, Xiao Yuan, Yang Qi, Zhenduo Wang, and Peng-Cheng Xie for insightful discussions. The authors are supported by the National Key R\&D Program of China
(No. 2021YFA1401900) and the National Science Foundation of China (No. 12174008). The computation was supported by High-performance Computing Platform of Peking University. Source code is available at https://github.com/PeilinZHENG/MLQC.

\bibliographystyle{apsrev4-1-title}
\bibliography{refs}

\end{document}


\title{Supplemental Materials: Ground-state properties via machine learning quantum constraints}

\author{Pei-Lin Zheng\textsuperscript{\tiny $\#$}}
\affiliation{International Center for Quantum Materials, Peking University, Beijing, 100871, China}
\affiliation{School of Physics, Peking University, Beijing, 100871, China}
\author{Si-Jing Du\textsuperscript{\tiny $\#$}}
\affiliation{International Center for Quantum Materials, Peking University, Beijing, 100871, China}
\affiliation{School of Physics, Peking University, Beijing, 100871, China}
\author{Yi Zhang}
\email{frankzhangyi@gmail.com}
\thanks{\textsuperscript{\tiny $\#$} P.-L. Zheng and S.-J. Du are responsible for 1D fermion chains and spin chains respectively and contributed equally.}
\affiliation{International Center for Quantum Materials, Peking University, Beijing, 100871, China}
\affiliation{School of Physics, Peking University, Beijing, 100871, China}

\maketitle

\section{Details on sample generation, supervised machine learning, and constrained optimization} \label{sec:sml}

Our strategy consists of three vital steps: sample generation ($\langle {\bf \hat O}\rangle$ and corresponding penalty), supervised machine learning to train ANNs on quantum constraints $f^{*}(\langle {\bf \hat O}\rangle)$, and constrained optimization upon target model systems.

\subsection{Sample generation}

In the main text, we propose two different paradigms of quantum constraints $f^{*}(\langle {\bf \hat O}\rangle)$: first, we can express the quantum constraints as an explicit function between the operator expectation values, e.g., $\langle \hat{O}_0\rangle=f^{*}(\langle \hat{O}_1\rangle,\ \cdots,\ \langle \hat{O}_N\rangle)$ for 1D Fermi seas. Here, we only need the physical operator expectation values $\langle {\bf \hat O}\rangle$ fully consistent with quantum constraints, which are straightforward to generate by selecting the appropriate ansatz and sampling the corresponding quantum many-body states.

On the other hand, in more general scenarios where relations between $\langle {\bf \hat O}\rangle$ are complex and implicit, we have to resort to a penalty function $f^{*}(\langle {\bf \hat O}\rangle)$ to characterize quantum constraints. Here, in addition to the physical $\langle {\bf \hat O}\rangle$ with $0$ penalty values, we generate unphysical $\langle {\bf \hat O}\rangle'$ by adding random small deviations $\delta \langle {\bf \hat O}\rangle$ to physical $\langle {\bf \hat O}\rangle$. We leave the discussions of the penalty values for such near-physical samples to the next subsection. 

We also exploit the equivalence through gauge transformations, which leave quantum constraints intact, for further simplification. Therefore, it suffices to keep one $\langle {\bf \hat O}\rangle$ example for each gauge equivalent class and reduces the degrees of freedom and thus the resulting computational complexity. For instance, for 1D fermion insulators with a bipartite unit cell, we can define the penalty function $f^{*}(g_r, \tilde g_{r'})$ for real $\tilde g_{\pm 1/2}$ only, and the penalty function for a generic $\{g_r, \tilde g_{r'}\}$ with complex $\tilde g_{\pm 1/2}$ is obtainable straightforwardly through a gauge transformation.

For reference, we show the pseudo-codes for sample generation of 1D fermion insulators with a bipartite unit cell and 1D interacting spin-1/2 chains in Algorithms \ref{alg:fc} and \ref{alg:sc}, respectively.

\begin{algorithm}
	\caption{Sample generation of 1D fermion insulators}
	\label{alg:fc}
	\begin{algorithmic}[1]
	    \Require
	    \Statex number $m$ of physical samples, ratio $p$ of unphysical to physical samples, truncation distance $\Lambda$, random decay rates $\alpha$
	    \Ensure
	    \Statex dataset $\bm{\mathcal{D}}$
	    \State $\bm{\mathcal{D}}=\{\}$;
		\For{$i=1$ to $m$}:
		    \State Generate a random vector $\bm{g'}_i=\{g'_r,\tilde g'_{r'}\}$;
		    \State $\bm{g'}_i \cdot =\{1|_{r=0},e^{-\alpha(|r|-1)}|_{r\ne0},e^{-\alpha(|r'|-0.5)}\}$ for locality;
		    \State Fourier transform $\bm{g'}_i$ to get $\bm{u'}_i=\{u'_k,\tilde{u}'_k\}$;
		    \State Normalize $\bm{u'}_i$ to get Bloch state $\bm{u}_i=\{u_k,\tilde{u}_k\}$;
		    \State Inverse Fourier transform $\bm{u}_i$ to get $\bm{g}_i=\{g_r,\tilde g_{r'}\}$;
		    \State Gauge transform $\bm{g}_i$ to make $\tilde g_{\pm 1/2}$ real;
		    \State $\bm{\mathcal{D}} \gets \{\bm{g}_i,\text{penalty value}=0\} $;
		    \State Run Algorithm \ref{alg:KLD} for basis vectors of physical manifold 
		    \State and return $\{\bm{g}_i+\delta\bm{g}_i^l, y^l_{KLD}\}_{l=1,\cdots,p}$;
	        \For{$l=1$ to $p$}:
		        \State $\bm{\mathcal{D}} \gets \{\bm{g}_i+\delta\bm{g}_i^l,\text{penalty value}=y^l_{KLD}\}$;
		    \EndFor
		\EndFor
	\end{algorithmic}
\end{algorithm}

\begin{algorithm}
	\caption{Sample generation of 1D interacting spin-1/2 chains}
	\label{alg:sc}
	\begin{algorithmic}[1]
	    \Require
	    \Statex number $m$ of physical samples, the ratio $p$ of unphysical to physical samples, truncation distance $l_{max}$, iMPS matrices dimension $\chi$
	    \Ensure
	    \Statex dataset $\bm{\mathcal{D}}$
	    \State $\bm{\mathcal{D}}=\{\}$;
		\For{$i=1$ to $m$}:
		    \State Generate random $\chi \times \chi$ matrices for iMPS $|\Psi_i \rangle$;
		    \State $\langle {\bf \hat O}\rangle_i =\langle\Psi_i|{\bf \hat O}|\Psi_i\rangle=\{S^\lambda_r,S^\lambda_r S^{\lambda'}_{r+l}|^{\lambda,\lambda'=x,y,z}_{l=1,\cdots,l_{max}-1}\}_{i}$;
		    \State $\bm{\mathcal{D}} \gets \{\langle {\bf \hat O}\rangle_i, \text{penalty value}=0\}$;
		    \State Run Algorithm \ref{alg:MSE} for basis vectors of physical manifold
		    \State and return $\{\langle{\bf \hat O}\rangle_i+\delta\langle{\bf \hat O}\rangle_i^l,y^l_{MSE}\}_{l=1,\cdots,p}$;
		    \For{$l=1$ to $p$}:
		        \State $\bm{\mathcal{D}} \gets \{\langle {\bf \hat O}\rangle_i+\delta\langle{\bf \hat O}\rangle_i^l,\text{penalty value}=y^l_{MSE}\}$;
		    \EndFor
		\EndFor
	\end{algorithmic}
\end{algorithm}

\subsection{Penalty values for near-physical samples}

We will employ two methods to define and evaluate the penalty value of near-physical (unphysical) samples $\langle {\bf \hat O}\rangle'=\langle {\bf \hat O}\rangle + \delta\langle {\bf \hat O}\rangle=\{\hat{O}_1,\hat{O}_2,\cdots\}$. Using the (weighted) norm of $\delta\langle {\bf \hat O}\rangle$ is an overestimate to the penalty value, as $\delta\langle {\bf \hat O}\rangle$ is not fully normal to the physical manifold in general, and we need to separate its projection along the physical manifold for a more sound estimation. 

Our first method starts from a Kullback-Leibler (KL) divergence perspective \cite{KLDiv, PME2022}. For such, we re-express the observables $\{\hat O_j\}$ as a set of two-level projection operators $\{\hat P_\gamma\}$, and correspondingly, the expectation values in terms of $\langle {\bf \hat P}\rangle=\{f_1,1-f_1,f_2,1-f_2,\cdots\}$ for a physical state, $\ f_\gamma=\langle\Phi_0|\hat P_\gamma|\Phi_0\rangle$, and $\langle {\bf \hat P}\rangle'=\{f_1',1-f_1',f_2',1-f_2',\cdots\}$ for an unphysical state, $f_\gamma'=f_\gamma+\delta f_\gamma$. Then, we can define the penalty value $y_{KLD}$ as the minimum total KL divergence between the given probabilities $\{f_\gamma'\} \in [0, 1]$ and the quantum probabilities $\langle\Phi|\hat P_\gamma|\Phi\rangle$ among all states:
\begin{eqnarray}
y_{KLD}&=&\min_{|\Phi\rangle}\sum_{\gamma}\left[f_\gamma'\log\left(\frac{f_\gamma'}{\langle\Phi|\hat P_\gamma|\Phi\rangle}\right)\right . \label{eq:PMEapp}\\& &\indent \indent \indent\ \left .+(1-f_\gamma')\log\left(\frac{1-f_\gamma'}{\langle\Phi|1-\hat P_\gamma|\Phi\rangle}\right)\right]  \nonumber\\
&\approx&\min_{|\Phi\rangle}\sum_{\gamma}\frac{\left(\delta f_\gamma-\langle\Phi_0|\hat P_\gamma|\delta\Phi\rangle-\langle\delta\Phi|\hat P_\gamma|\Phi_0\rangle\right)^2}{2f_\gamma(1-f_\gamma)}, \nonumber 
\end{eqnarray}
where we have assumed $\delta f_\gamma$ is small and $|\Phi\rangle$ is not far from $|\Phi_0\rangle$, and performed a second-order Taylor expansion to the last step:
\begin{equation}
|\Phi\rangle=\frac{|\Phi_0\rangle+|\delta\Phi\rangle}{\sqrt{(\langle\Phi_0|+\langle\delta\Phi|)(|\Phi_0\rangle+|\delta\Phi\rangle)}}\ . \label{eq:Phi}
\end{equation} 
$y_{KLD}$ is non-negative-definite. $y_{KLD}=0$ if and only if $\{f_\gamma'\}$ is physical with an existing $|\Phi\rangle$ so that $\{f_\gamma'\} = \langle\Phi|\hat P_\gamma|\Phi\rangle$; otherwise, $y_{KLD}>0$ and describes the extent to which $\{f_\gamma'\}$ deviates from the physical manifold, measured from the closest quantum state $|\Phi\rangle$ towards $\{f_\gamma'\}$ with the best effort. 

\begin{figure}
	\includegraphics[width=1.0\linewidth]{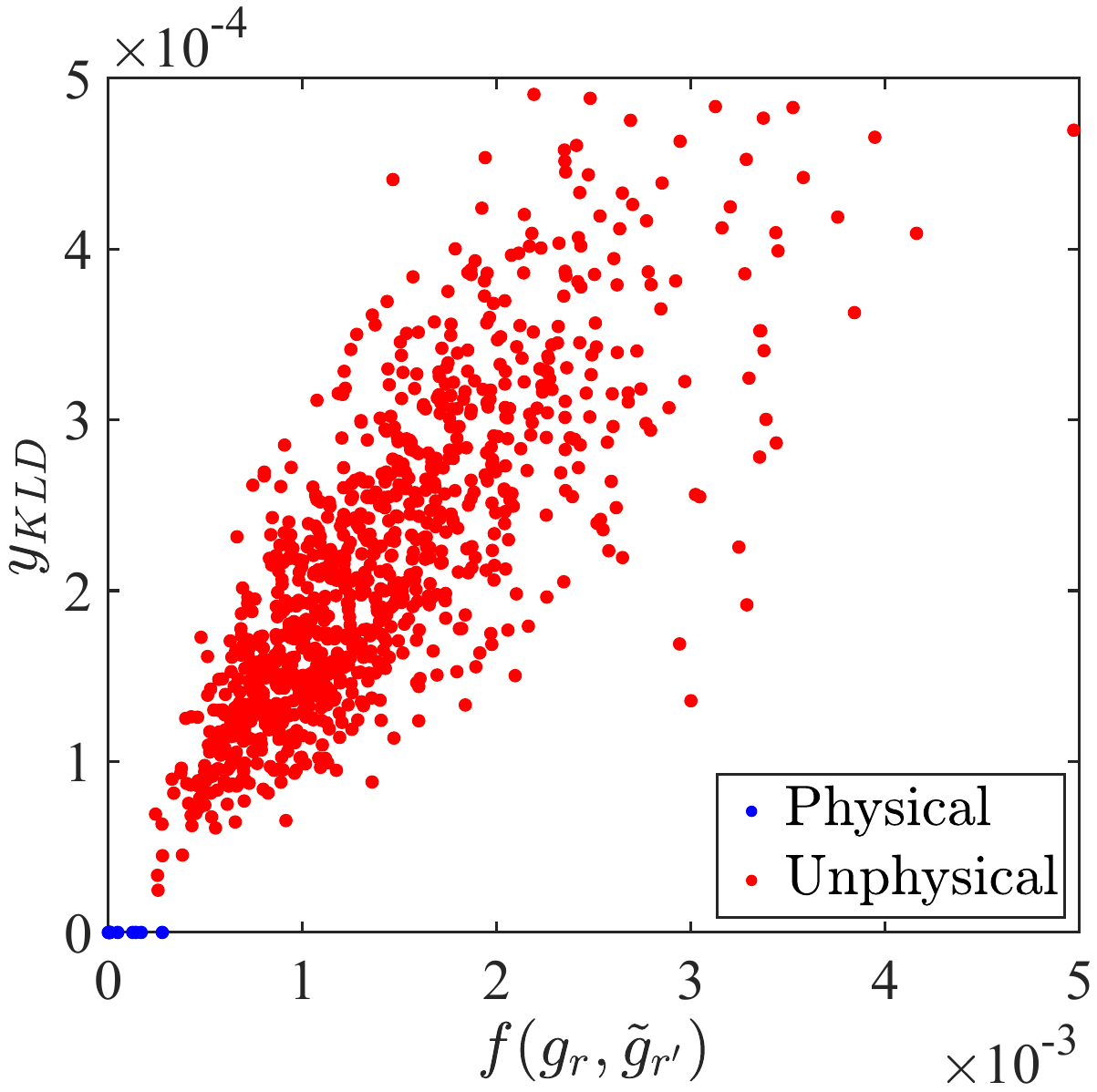}
	\caption{The $y_{KLD}$ in Eq. \ref{eq:PMEapp} and the benchmark $f(g_r,\tilde{g}_{r'})$ in Eq. 7 in the main text show a positive connection and good consistency between them.} \label{fig:PMEins}
\end{figure}

In practice, we first sample $|\delta\Phi\rangle$ and perform a principal component analysis (PCA) on $\{\langle\Phi_0|\hat{\bm P}|\delta\Phi\rangle+\langle\delta\Phi|\hat{\bm P}|\Phi_0\rangle\}$ - the basis vectors in the physical manifold; then, we project out these directions from $\{\delta f_\gamma\}$, and evaluate the penalty value $y_{KLD}$ following Eq. \ref{eq:PMEapp}. We summarize the pseudo-code in Algorithm \ref{alg:KLD}. In Fig. \ref{fig:PMEins}, we show a comparison between $y_{KLD}$ and the benchmark $f(g_r,\tilde{g}_{r'})$ following Eq. 7 in the main text. We observe a positive connecting between the two penalty functions, suggesting that $y_{KLD}$ is a generally available and reasonably consistent indicator for quantum constraints. 

\begin{algorithm}
	\caption{Evaluation of penalty value $y_{KLD}$}
	\label{alg:KLD}
	\begin{algorithmic}[1]
	    \Require
	    \Statex number $p$ of unphysical state per physical state $|\Phi_0\rangle$, the corresponding physical sample $\bm{g}$, number $n$ of reference quantum-state samples, truncation threshold $\epsilon$
	    \Ensure
	    \Statex $\{\bm{g}+\delta\bm{g}^l, y^l_{KLD}\}_{l=1,\cdots,p}$
	    \State Construct $\hat {\bm P}=\{\hat P_\gamma\}$ from the operators of $\bm{g}$;
	    \State $\bm{X}=[]$;
		\For{$i=1$ to $n$}:
		    \State Generate random $|\Psi_i\rangle$ that satisfies $\langle\Phi_0|\Psi_i\rangle=0$;
		    \State Compute $\bm{x}_i= \langle \Phi_0|\hat {\bm P}|\Psi_i\rangle+\langle\Psi_i|\hat {\bm P}|\Phi_0\rangle$;
		    \State $\bm{X} \gets$ normalized $\bm{x}_i$;
		 \EndFor   
		 \State Compute covariance matrix $\bm{K}_{\bm{XX}}=(\bm{X}- \bm{\bar{X}})(\bm{X}^T - \bm{\bar{X}}^T)$;
		 \State Diagonalize $\bm{K}_{\bm{XX}}=\bm{V}\bm{A}\bm{V}^{T}$;
		 \State Drop eigenvectors with eigenvalues less than $\epsilon$ from $\bm{V}$;
		 \State Get the projector to physical manifold $\bm{W}=\bm{V}\bm{V}^T$;
		 \For{$l=1$ to $p$}:
  	         \State Generate a random vector $\delta\bm{g}^l=\{\delta g_r,\delta\tilde{g}_{r'}\}$;
  	         \State Compute $\{f_\gamma\}$ and $\delta\bm{P}^l$ from $\bm{g}$ and $\delta\bm{g}^l$;
    		 \State Calculate $(\bm{I}-\bm{W}) \delta\bm{P}^l$ then $y_{KLD}^l$ following Eq. \ref{eq:PMEapp};
    	 \EndFor 
	\end{algorithmic}
\end{algorithm}

Our second method is based upon the mean squared error (MSE). It defines the penalty value for expectation values $\langle {\bf \hat O}\rangle'$ as the following expression:
\begin{eqnarray}
y_{MSE}&=&\min_{|\Phi\rangle}\sum_{j}\left(\langle \hat{O}_j\rangle'-\langle\Phi|\hat O_j|\Phi\rangle\right)^2 \nonumber \\
&=&\min_{|\Phi\rangle}\sum_{j}\left[\delta \langle \hat{O}_j \rangle-\left(\langle\Phi|\hat O_j|\Phi\rangle-\langle \hat{O}_j \rangle\right)\right]^2. 
\label{eq:MSEapp}
\end{eqnarray}
Similar to $y_{KLD}$, for small deviations $\delta \langle \hat{O}_j \rangle$ and $|\Phi\rangle$ similar to Eq. \ref{eq:Phi}, we can sample $\langle\Phi|\bf \hat{O}|\Phi\rangle-\langle \bf \hat{O}\rangle$ for basis vectors in the physical manifold, and evaluate $y_{MSE}$ following Eq. \ref{eq:MSEapp}. We summarize the corresponding pseudo-code in Algorithm \ref{alg:MSE}.

\begin{algorithm}
	\caption{Evaluation of penalty value $y_{MSE}$}
	\label{alg:MSE}
	\begin{algorithmic}[1]
	    \Require
	    \Statex the set of observables $\{\hat O_j\}$, number $p$ of unphysical state per physical state $|\Phi_0\rangle$, the corresponding physical sample $\langle {\bf \hat O}\rangle$, number $n$ of reference quantum-state samples, truncation threshold $\epsilon$, hyper-parameter $\delta$
	    \Ensure
	    \Statex $\{\langle{\bf \hat O}\rangle+\delta\langle{\bf \hat O}\rangle^l,y^l_{MSE}\}_{l=1,\cdots,p}$
	    \State $\bm{X}=[]$;
		\For{$i=1$ to $n$}:
		    \State Randomly generate $|\Psi_i\rangle$;
		    \State Compute $|\Phi\rangle=\frac{|\Phi_0\rangle+\delta|\Psi_i\rangle}{\sqrt{(\langle\Phi_0|+\delta\langle\Psi_i|)(|\Phi_0\rangle+\delta|\Psi_i\rangle)}}$;
		    \State Compute $\bm{x}_i=\left[\langle\Phi|\hat O_j|\Phi\rangle-O_j^0\right]_{j=1,2,\cdots}$;
		    \State Normalize $\bm{x}_i$;
		    \State $\bm{X} \gets \bm{x}_i$;
		 \EndFor   
		 \State Compute covariance matrix $\bm{K}_{\bm{XX}}=(\bm{X}- \bm{\bar{X}})(\bm{X}^T - \bm{\bar{X}}^T)$;
		 \State Diagonalize $\bm{K}_{\bm{XX}}=\bm{V}\bm{A}\bm{V}^{T}$;
		 \State Drop eigenvectors with eigenvalues less than $\epsilon$ from $\bm{V}$;
		  \State Get the projector to physical manifold $\bm{W}=\bm{V}\bm{V}^T$;
		 \For{$l=1$ to $p$}
		     \State Generate a random vector $\delta\langle{\bf \hat O}\rangle^l=\{\delta \langle \hat{O}_j \rangle\}$;
		     \State Calculate $(\bm{I}-\bm{W})\delta\langle{\bf \hat O}\rangle^l$ then $y_{MSE}^l$ following Eq. \ref{eq:MSEapp};
    	 \EndFor 
	\end{algorithmic}
\end{algorithm}

In case we have not exhausted the basis vectors of the physical manifold, the evaluated penalty values may have discrepancies with the definitions in Eqs. \ref{eq:PMEapp} and \ref{eq:MSEapp}. In the extreme case, unphysical samples with finite penalty values may be physical. Fortunately, the probability of such occurrence should be small; otherwise, it may cause the penalty function $f^{*}(\langle {\bf \hat O}\rangle)$ to flatten and the restoring force toward the physical expectation values to vanish. Such issues, commonly known as the barren plateau, can be fixed by further training with designated samples and other newly developed solutions \cite{RolnickVBS17, KimYYK19, 8850096} designated for machine learning with potentially noisy data \cite{BrodleyF99, BeigmanK09}.

\subsection{Artificial neural networks and supervised machine learning}

\begin{figure}
	\includegraphics[width=0.75\linewidth]{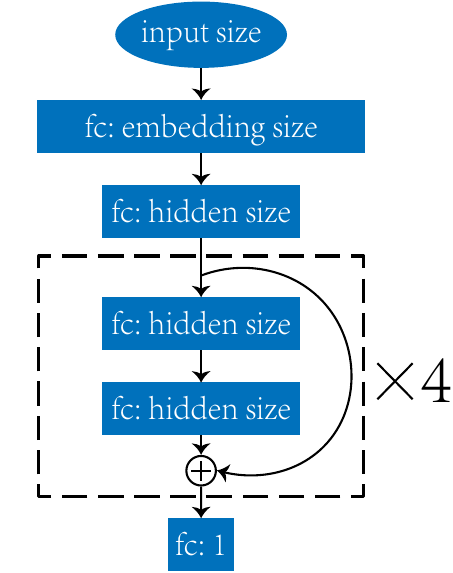}
	\caption{The architecture of our ANNs used as the $f^{*}$ function for the quantum constraints consists of a couple of fully connected (fc) layers and four residual blocks with potential shortcuts. Our choices for the ANN layer width (number of neurons in a layer) vary with the target systems: we set embedding size and hidden size as 500 and 100 for the 1D fermion chains and 600 and 144 for the 1D spin chains, respectively.} \label{fig:NNarch}
\end{figure}

We apply supervised machine learning on the generated datasets to train our ANNs, each of which consists of two hidden layers and four residual blocks \cite{He2016resnet} with ReLU activation functions \cite{Fuku1969relu, glorot2011relu} and batch normalization \cite{Ioffe2015bn}; see Fig. \ref{fig:NNarch} for an illustration of our ANN architecture. The output neuron has a sigmoid activation function \cite{Han1995sigmoid} and (its average over multiple ANNs) stands for the $f^*$ function on the quantum constraints.

For 1D fermion chains, we first apply the function $\tanh(500y)$ to stretch the label $y$ for better compatibility with the range $[0,1]$ of the sigmoid function of the ANN output neuron. We use the mean-square error as our cost function and apply the stochastic gradient descent method with the Adam optimizer with a learning rate of $10^{-3}$, L2 regularization of $10^{-6}$, a batch size of 4096, and the StepLR scheduler with a learning rate decay ratio of 0.5 for every 200 epochs. We train multiple ANNs in parallel for 1000 epochs to guarantee good convergence. For each ANN, we keep the best performing one on a separate validation set during its entire training process, which takes less than 12 hrs on one NVIDIA RTX 3090 GPU.

For 1D spin-1/2 chains, we stretch the label $y$ with the function $\tanh{y}$ for better regression results. We use the mean-square error as the cost function and apply the stochastic gradient descent method with the Adam optimizer with an initial learning rate of $10^{-3}$ and L2 regularization of $10^{-5}$, and a batch size of 360. We also employ the StepLR scheduler with a multiplier of 0.5 to the learning rate for every 50 epochs. We train multiple ANNs in parallel for at least 500 epochs to ensure good convergence and keep the ANN with the minimum loss on a validation set separated from the training set. The entire training process takes less than 12 hrs on one NVIDIA RTX 3090 GPU.

\subsection{Constrained optimization}

With the quantum constraints $f$ or $f^*$, we employ an adiabatic-algorithm-based constrained optimization algorithm for the ground-state properties of target Hamiltonians or design targets, which we summarize in Algorithm \ref{alg:opt}. Our choices of sample datasets concentrate on scenarios consistent or nearly consistent with the quantum constraints, so we need to avoid deeply unphysical scenarios where our ANNs have little or no experience during their supervised machine learning and do not respond meaningfully. Therefore, we recommend starting with initial conditions $\langle{\bf \hat O}\rangle_i$ that are known to be consistent with the quantum constraints, either via exact solutions or controlled approximation, and then change the optimization targets gradually, using each step's optimal $\langle{\bf \hat O}\rangle_0$ as the initial condition for the next, and so on so forth. For example, as we tune the parameter of a Hamiltonian in small steps, the ground-state properties $\langle{\bf \hat O}\rangle$ of one step should not be too far from its neighboring steps, except for the case of an abrupt first-order phase transition. Such a scheme keeps our searches in the `river bed' of physical or near physical $\langle{\bf \hat O}\rangle$ that the ANNs are accustomed to.

\begin{figure}
	\includegraphics[width=1.0\linewidth]{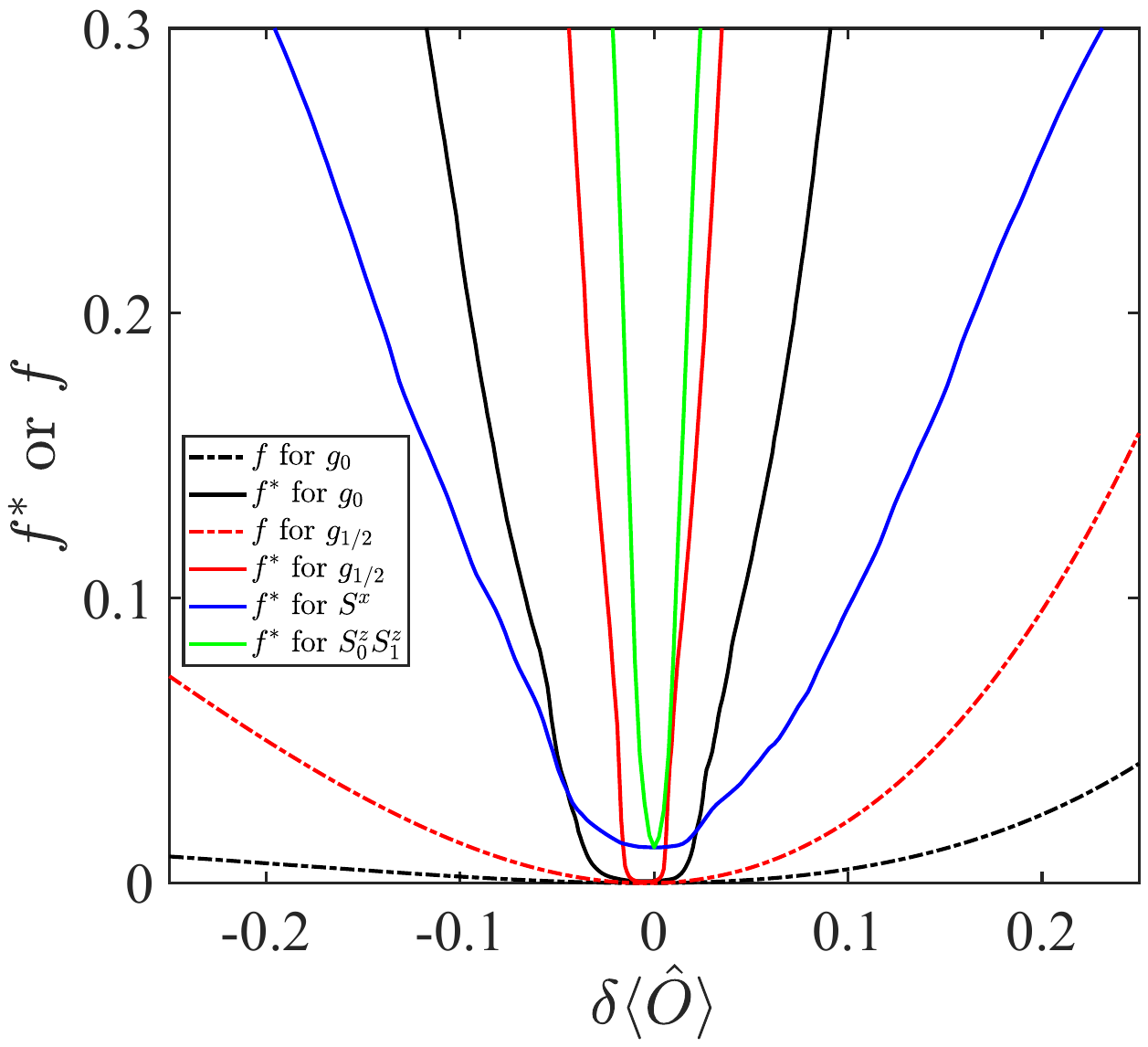}
	\caption{The penalty functions with respect to selected expectation values' departure from the quantum constraints illustrate their tendency to restore the expectation values to physically allowed values. We note that the strengths of the quantum constraints may vary from case to case, e.g., the ANN $f^{*}$ and the benchmark $f$ for departure of $g_0$ and $\tilde{g}_{1/2}$ in 1D fermion insulators, and the ANN $f^{*}$ for departure of $\langle S^x_0\rangle$ and $\langle S^z_0 S^z_1\rangle$ in 1D spin chains. } \label{fig:restore}
\end{figure}

\begin{algorithm}
	\caption{Constrained Optimization}
	\label{alg:opt}
	\begin{algorithmic}[1]
	    \Require
	    \Statex tuning parameter $\bm{t}$, target function $h(\langle{\bf \hat O}\rangle; t)$, quantum constraints $f^{*}(\langle{\bf \hat O}\rangle)$, initial $\langle{\bf \hat O}\rangle_i$, penalty weight $\eta$
	    \Ensure
	    \Statex $\bm{h_0}(\bm{t})=\mathop{\min}_{\langle {\bf \hat O}\rangle} [\bm{h}(\bm{t})],\ \bm{O_0}(\bm{t})=\mathop{\arg\min}_{\langle {\bf \hat O}\rangle}[\bm{h}(\bm{t})]$
		\State $\bm{h_0}(\bm{t})=\{\},\ \bm{O_0}(\bm{t})=\{\}$;
		\For{$t$ in $\bm{t}$}:
		    \State $L(\langle{\bf \hat O}\rangle)=h(\langle{\bf \hat O}\rangle; t)+\eta f^{*}(\langle{\bf \hat O}\rangle)$;
            \State Minimize $L$ by gradient-based algorithms with $\langle {\bf \hat O}\rangle_i$ as \State the initial point for $h_0$ and $\langle {\bf \hat O}\rangle_0$;
		    \State $\bm{O_0}(\bm{t}) \gets \langle {\bf \hat O}\rangle_0=\arg\min_{\langle{\bf \hat O}\rangle\in\bm{S^+}} [h(t)]$;
		    \State $\bm{h_0}(\bm{t}) \gets h_0 =h(\langle{\bf \hat O}\rangle_0; t) $;
		    \State $\langle {\bf \hat O}\rangle_i=\langle {\bf \hat O}\rangle_0$;
		\EndFor
	\end{algorithmic}
\end{algorithm}

The rough landscape of an ANN may cause additional complications, and trap the optimization in local minimums, or mislabel unphysical $\langle{\bf \hat O}\rangle$ as physical and cause a diversion or even a mistake. To counter these issues, we employ the idea of ensemble learning and use the average output of multiple ANNs as the value of $f^{*}(\langle {\bf \hat O}\rangle)$. We also set a threshold that the maximum outputs of multiple ANNs should not exceed to eliminate the likelihood of false `physical,' and keep multiple searches in parallel and multiple $\langle{\bf \hat O}\rangle$ outcomes as the seeds for the next step to increase the acceptance rate. In addition, we include small, random deviations from each seed before optimization to avoid being trapped in local minimums. We note that these tricks are no longer necessary when we use the well-behaved quantum-constraint expressions such as $f$, which only requires a few ($O(1)$) initial points for the search to avoid cases of falling into a local minimum. Intuitively, the analytical function $f$ is much smoother than $f^*$, as illustrated in Fig. \ref{fig:restore}. At the same time, the target physical region in the $\langle{\bf \hat O}\rangle$ space itself should not cause too much roughness concern as the expectation values evolve gradually with the underlying quantum many-body states. This suggests there is certainly room for improvement on the approximation quality and the landscape roughness of the ANNs.

In practice, we parallelize multiple searches from different seeds (initial conditions) and use the Adam optimizer with a step size of $10^{-3}$ and the StepLR scheduler with a slow-down ratio of $0.5$ for every 200 steps for an overall 800 steps on target minimization. We note that Algorithm \ref{alg:opt} uses only gradient information, and in principle, adding the information of the Hessian matrix can yield better results, which we leave to future studies.

In addition, we can also use a classification ANN as $f^*$ that labels physical (unphysical) $\langle{\bf \hat O}\rangle$ samples as 0 (1). As the landscape of such $f^*$ will have a large number of barren plateaus and steep cliffs, simple gradient-based optimizations are no longer suitable, and non-derivative optimization methods, such as genetic algorithm, optimization by quadratic approximation \cite{Powell1994, Powell2002}, may be considered in the future. However, we note that the non-convex optimization problem is NP-hard, and there is no guarantee that the result obtained must be a global minimum irrespective of the optimization method.

\section{Machine learning quantum constraints for 1D Fermi seas}

\subsection{Quantum constraints for multiple Fermi seas}

Generalizing the 1D Fermi sea scenarios to ground states with multiple Fermi seas, we can identify a ground state with the set of Fermi points $\{k_n,n=1,\cdots,2n_{fs}\}$, arranged from small to large, where $n_{fs}$ is an upper bound of the number of Fermi seas \footnote{Fewer Fermi seas can be represented as identical values in $\{k_n\}$.}. The vital expectation values for the ground state are its two-point correlators $C_r = \langle c^\dagger_{x+r}c_{x}\rangle$:
	\begin{eqnarray}
	\label{cor}
	C_0 &=& \frac{\sum_{n=1}^{2n_{fs}}(-1)^n k_n}{2\pi} \label{eq:1dfs_Cr} \\ \nonumber
	C_r &=& \frac{\sum_{n=1}^{2n_{fs}}(-1)^n \exp(\text{i}k_n r)}{2\pi\text{i}r},\ r \ge 1,
	\end{eqnarray}
which are complex-value decaying functions of the correlation distance $r$. By sampling $\{k_n,n=1, \cdots, 2n_{fs}\}$, we obtain our expectation-value dataset following Eq. \ref{eq:1dfs_Cr}. Then, with $\left\{\frac{C_1}{C_0},\cdots,\frac{C_{n_{fs}}}{C_0}\right\}$ as the inputs and $C_0$ as the intended output, we train our ANNs on a quantum constraint in the form of $C_0=f^{*}\left(\frac{C_1}{C_0},\cdots,\frac{C_{n_{fs}}}{C_0}\right)$ via supervised machine learning. It is straightforward to see from Eq. \ref{eq:1dfs_Cr} that such $f^*$ is a well-defined single-valued function.

For demonstration, we consider the $n_{fs}=3$ case, and prepare a dataset of $7\times10^{5}$ samples in the training set and $1.4\times10^{5}$ samples in the validation set. Through supervised machine learning detailed in the previous section, the ANNs show good generalizability for the quantum constraints of 1, 2, or 3 Fermi seas. We also cross-check training sets of $5\times10^{5}$ samples with exactly 1, 2, or 3 Fermi seas on validation sets of $1\times10^{5}$ samples. Unsurprisingly, the ANNs trained with more Fermi seas generalizes well to the cases with fewer Fermi seas, but not vice versa. We use the average output of two ANNs trained on operator expectation values from 1, 2, and 3 Fermi seas and two ANNs trained on exactly 3 Fermi seas as the quantum constraints $f^{*}$ for the subsequent constrained optimization processes, whose details are discussed in the previous section.

\begin{figure}
		\includegraphics[width=1.0\linewidth]{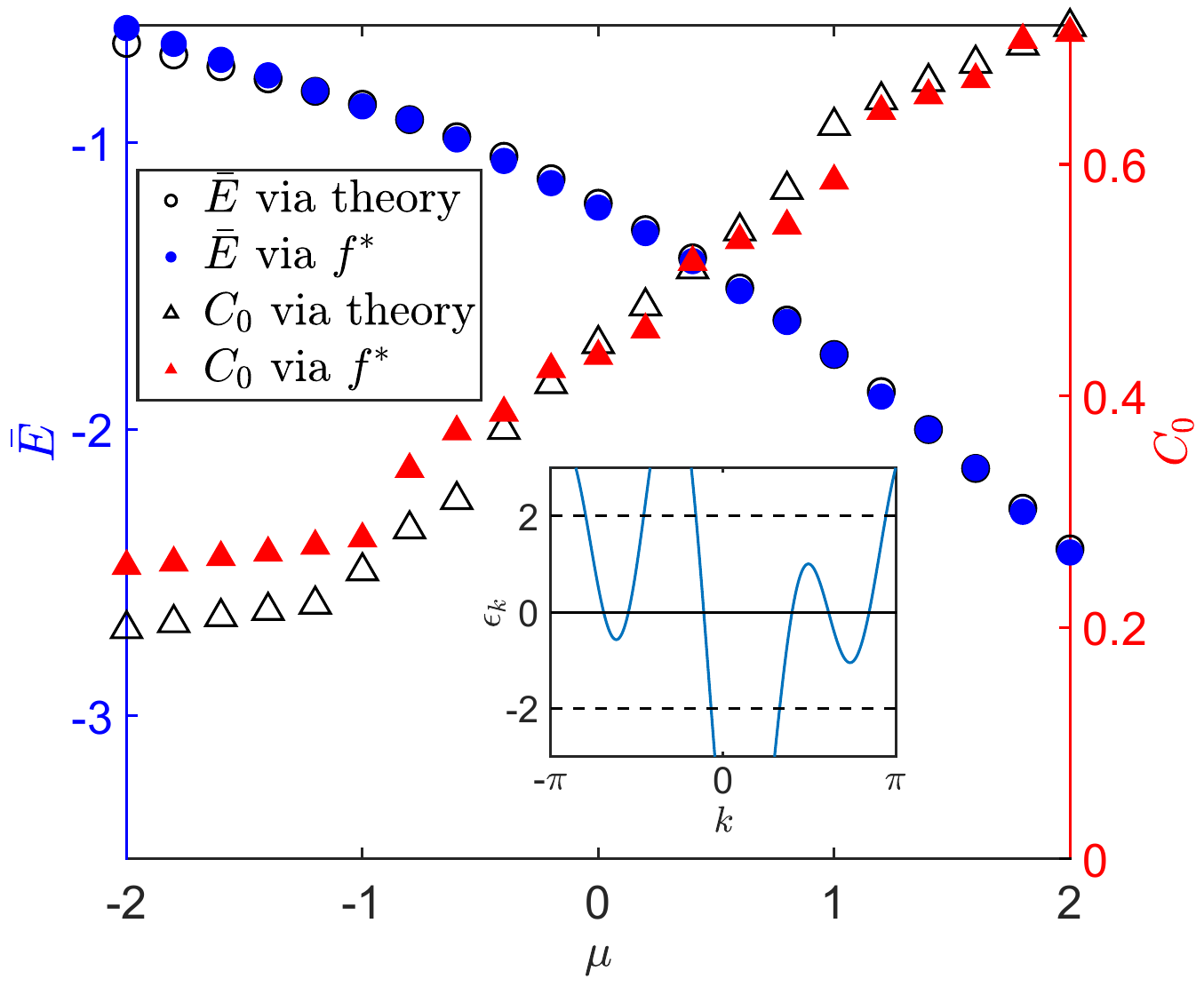}
		\caption{The expectation values of the ground-state energy $\bar{E}$ and electron density $C_0$ for the Hamiltonian in Eq. \ref{eq:Ham} versus model parameter $\mu$ show good consistency between conventional benchmarks and our approach via quantum constraints with ANN $f^{*}$. $t_1=1-\text{i}$, $t_2=0.5-0.9\text{i}$, $t_3=1-\text{i}$. Inset: the dispersion $\epsilon_k$ in Eq. \ref{eq:1dmetaldisp} for the same model. The number of Fermi seas alters between one and three for Fermi energy $\mu\in [-2, 2]$ between the dashed lines.} \label{fig:metal_result}
\end{figure}

We can apply the obtained $f^{*}$ towards the ground-state properties of the following Hamiltonians:
	\begin{eqnarray}
	\hat{H}=\sum_{x}\left[-\sum_{r=1}^{r_{cut}}\left(t_r c_{x+r}^{\dagger}c_{x}+\text{h.c.}\right) -\mu c_{x}^{\dagger}c_{x} \right],   \label{eq:Ham}
	\end{eqnarray}
where $t_r$ is the complex $r^{th}$-nearest-neighbor hopping amplitude, $1\le r\le r_{cut}$, and $\mu$ is the Fermi energy. We can represent the average energy per site as the operator expectation values in Eq. \ref{cor}:
	\begin{eqnarray}
	\bar{E}&=&\left\langle H\right\rangle/N=-\mu C_0-\sum_{r=1}^{r_{cut}} t_r C_r+\text{c.c.} \label{eq:optim} \\
	&=& -\left[ \mu+ \sum_{r=1}^{r_{cut}} (t_r \frac{C_r}{C_0} +\mbox{c.c.}) \right] f^{*}\left(\frac{C_1}{C_0},\frac{C_2}{C_0},\cdots,\frac{C_{n_{fs}}}{C_0}\right), \nonumber
	\end{eqnarray}
where the introduction of $f^{*}$ in the second line requires $n_{fs} \le r_{cut}$.
In hindsight, the $r_{cut}^{th}$-nearest-neighbor hopping can potentially give rise to $r_{cut}$ Fermi seas in the ground states, see Fig. \ref{fig:metal_result} inset; therefore, the quantum-state samples should at least incorporate such scenarios for adequate diversity and generalizability.

Consequently, we transform the problem into a conventional optimization for the minimum of $\bar{E}$ with respect to the parameters $\left\{\frac{\mbox{Re}(C_1)}{C_0},\frac{\mbox{Im}(C_1)}{C_0},\cdots,\frac{\mbox{Re}(C_{fs})}{C_0},\frac{\mbox{Im}(C_{fs})}{C_0}\right\}$. Since there is no unphysical expectation-value settings to worry about during the optimization in this formalism, it suffices to start from a random initial condition consistent with $n_{fs}=3$, and minimize $\bar{E}$ in Eq. \ref{eq:optim} with Adam optimizer and StepLR scheduler for a given Hamiltonian in Eq. \ref{eq:Ham}. For instance, we summarize the results of the obtained expectation values of ground-state energy $\bar{E}$ and electron density $C_0$ for $t_1=1-\text{i}$, $t_2=0.5-0.9\text{i}$, $t_3=1-\text{i}$ and $\mu\in \left[-2,2\right]$ in Fig. \ref{fig:metal_result}.

On the other hand, we can solve the ground states of Eq. \ref{eq:Ham} in the momentum space $\hat{H}=\sum_k \left(\epsilon_k-\mu\right) c^\dagger_k c_k$,
\begin{eqnarray}
\epsilon_k = -2 \sum_{r=1}^{r_{cut}}\mbox{Re}(t_r)\cos(k r)-\mbox{Im}(t_r)\sin(k r),  \label{eq:1dmetaldisp}
\end{eqnarray}
see Fig. \ref{fig:metal_result} inset. For a given $\mu$, the ground state fill the Fermi sea $c^\dagger_k c_k=1$ for $\epsilon_k<\mu$ and $c^\dagger_k c_k=0$ otherwise, and we have up to $n_{fs}= 3$ Fermi seas for $\mu \in \left[-2,2\right]$. With these ground states, we can calculate the exact values of $\bar{E}$ and $C_0$ and use them as benchmarks in the Fig. \ref{fig:metal_result}. In most cases, our approach gets satisfactorily accurate results, where the average error of $\bar{E}$ and $C_0$ are $1.3\times10^{-2}$ and $2.8\times10^{-2}$, respectively.

\subsection{Additional details on machine learning quantum constraints for 1D fermion insulators}

In this subsection, we provide further details and results on our quantum-constraints strategy on 1D fermion insulators with a bipartite unit cell.

First, we note the following transformations between correlators and projection operators:
\begin{eqnarray}
\hat{P}_0^{AA(BB)}&=&\hat{c}_{i}^\dagger\hat{c}_{i}, \nonumber\\
\hat{P}_r^{AA(BB)}&=&\frac{\left(\hat{c}_{i+r}^\dagger+\hat{c}_{i}^\dagger\right)\left(\hat{c}_{i+r}+\hat{c}_{i}\right)}{2},r\in\mathbb{Z}^{+}, \nonumber\\
\hat{P'}_r^{AA(BB)}&=&\frac{\left(\hat{c}_{i+r}^\dagger+\text{i}\hat{c}_{i}^\dagger\right)\left(\hat{c}_{i+r}-\text{i}\hat{c}_{i}\right)}{2},r\in\mathbb{Z}^{+},\\
\hat{P}_{r'}^{AB}&=&\frac{\left(\hat{c}_{i+r'}^\dagger+\hat{c}_{i}^\dagger\right)\left(\hat{c}_{i+r'}+\hat{c}_{i}\right)}{2},r'\in\mathbb{Z}+\frac{1}{2}, \nonumber\\
\hat{P'}_{r'}^{AB}&=&\frac{\left(\hat{c}_{i+r'}^\dagger+\text{i}\hat{c}_{i}^\dagger\right)\left(\hat{c}_{i+r'}-\text{i}\hat{c}_{i}\right)}{2},r'\in\mathbb{Z}+\frac{1}{2}; \nonumber
\end{eqnarray}
and the corresponding expectation values are:
\begin{eqnarray}
\langle \hat P_0^{AA} \rangle&=&C_0^{AA}=\frac{1+g_0}{2}, \nonumber\\
\langle \hat P_0^{BB} \rangle&=&C_0^{BB}=\frac{1-g_0}{2}, \nonumber\\
\langle \hat P_r^{AA} \rangle&=&\mbox{Re}(C_r^{AA})+C_0^{AA}=\frac{\mbox{Re}(g_r)+1+g_0}{2}, \nonumber\\
\langle \hat P_r^{BB} \rangle&=&\mbox{Re}(C_r^{BB})+C_0^{BB}=\frac{-\mbox{Re}(g_r)+1-g_0}{2}, \nonumber\\
\langle {\hat{P'}}_r^{AA} \rangle&=&\mbox{Im}(C_r^{AA})+C_0^{AA}=\frac{\mbox{Im}(g_r)+1+g_0}{2},\\
\langle {\hat{P'}}_r^{BB} \rangle&=&\mbox{Im}(C_r^{BB})+C_0^{BB}=\frac{-\mbox{Im}(g_r)+1-g_0}{2}, \nonumber\\
\langle \hat P_{r'}^{AB} \rangle&=&\mbox{Re}(C_{r'}^{AB})+\frac{C_0^{AA}+C_0^{BB}}{2}=\frac{\mbox{Re}(\tilde{g}_{r'})+1}{2}, \nonumber\\
\langle {\hat{P'}}_{r'}^{AB} \rangle&=&\mbox{Im}(C_{r'}^{AB})+\frac{C_0^{AA}+C_0^{BB}}{2}=\frac{\mbox{Im}(\tilde{g}_{r'})+1}{2}. \nonumber
\end{eqnarray}

\begin{figure}
	\includegraphics[width=1.02\linewidth]{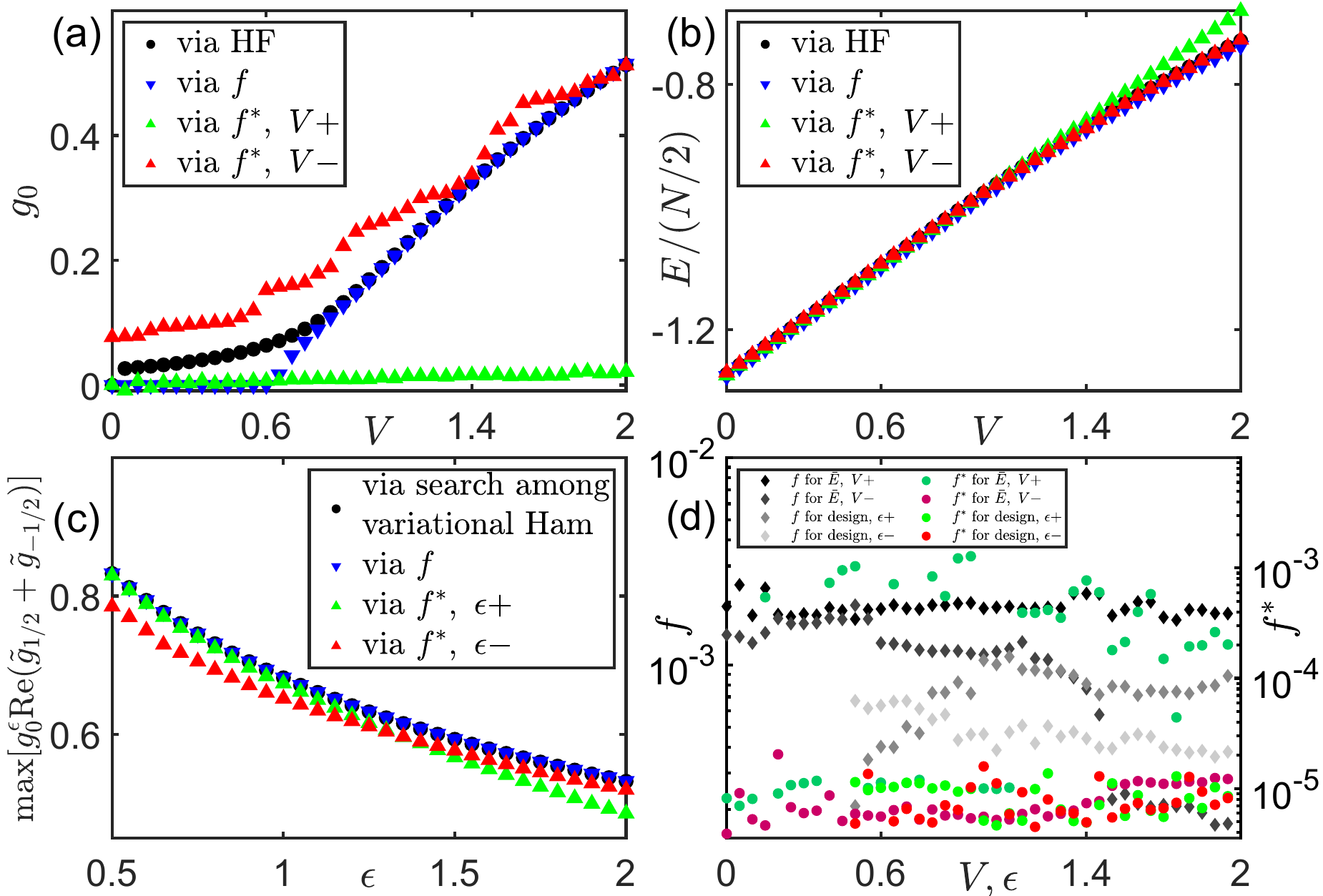}
	\caption{(a) The density-wave order parameters $g_0 = C_0^{AA} - C_0^{BB}=\langle c^\dagger_A c_A \rangle - \langle c^\dagger_B c_B \rangle$ of the mean-field ground states for the Hamiltonian in Eq. 8 in the main text show a consistent tendency of monotonic increase with respect to $V$ yet with considerable discrepancies between the Hartree-Fock method, the quantum-constraints approach with the theoretical expression $f$ and the ANNs $f^{*}$. For the latter, we consider two sweeps with increasing and decreasing $V$, where $\{g_r, \tilde g_{r'}\}$ are tracked and optimized step by step. (b) Similar to Fig. 2 top in the main text, the ground-state energy shows better consistency across the board of various approaches. Here we also show additional results from $f^{*}$ with increasing $V$ from $V=0$. (c) Similar to Fig. 3 in the main text, the maximum of $(\mbox{Re}(g_0))^\epsilon \mbox{Re}(\tilde g_{1/2} + \tilde g_{-1/2})$, instead of $(\mbox{Re}(g_1))^\epsilon \mbox{Re}(\tilde g_{1/2} + \tilde g_{-1/2})$, shows good consistency between constrained optimization via the quantum constraints $f$ and searches among the variational Hamiltonian, yet a discrepancy develops for values based upon the ANN quantum constraints $f^{*}$. (d) The ANN $f^{*}(g_r, \tilde g_{r'})$ and the benchmark $f^{*}(g_r, \tilde g_{r'})$ suggest that all $\{g_r, \tilde g_{r'}\}$ expectation values in the previous panels obey the quantum constraints well. } \label{fig:od}
\end{figure}

Moreover, we notice that while measurements of most expectation values such as the ground-state energies are precise and straightforward, the estimation for the order parameters of the spontaneous symmetry breaking phase is relatively crude and dependent on the quality of the quantum constraints and details of the constrained optimization, see Fig. \ref{fig:od} for illustrations. We attribute such difficulties to two main factors: the order parameter is relatively soft, and small deviations of it do not yield a meaningful rise in energy, making the constrained optimization relatively insensitive to its value; on the other hand, the ANNs $f^{*}$ tend to have lower penalties for deviations in the order parameter thus less restoring power to enforce consistency with the quantum constraints, see Fig. \ref{fig:restore}. Adding more specific training samples to emphasize the quantum constraints on the order parameter may improve the ANNs in such aspects, which we leave to future work. In comparison, we note that the order parameter from the high-quality quantum constraints $f$ is able to achieve satisfactory consistency. Its deviations at small $V$ from Hartree-Fock results are likely due to the impact of the cut-off $\Lambda$ as the mean-field gap diminishes.

\begin{figure}
	\includegraphics[width=1.0\linewidth]{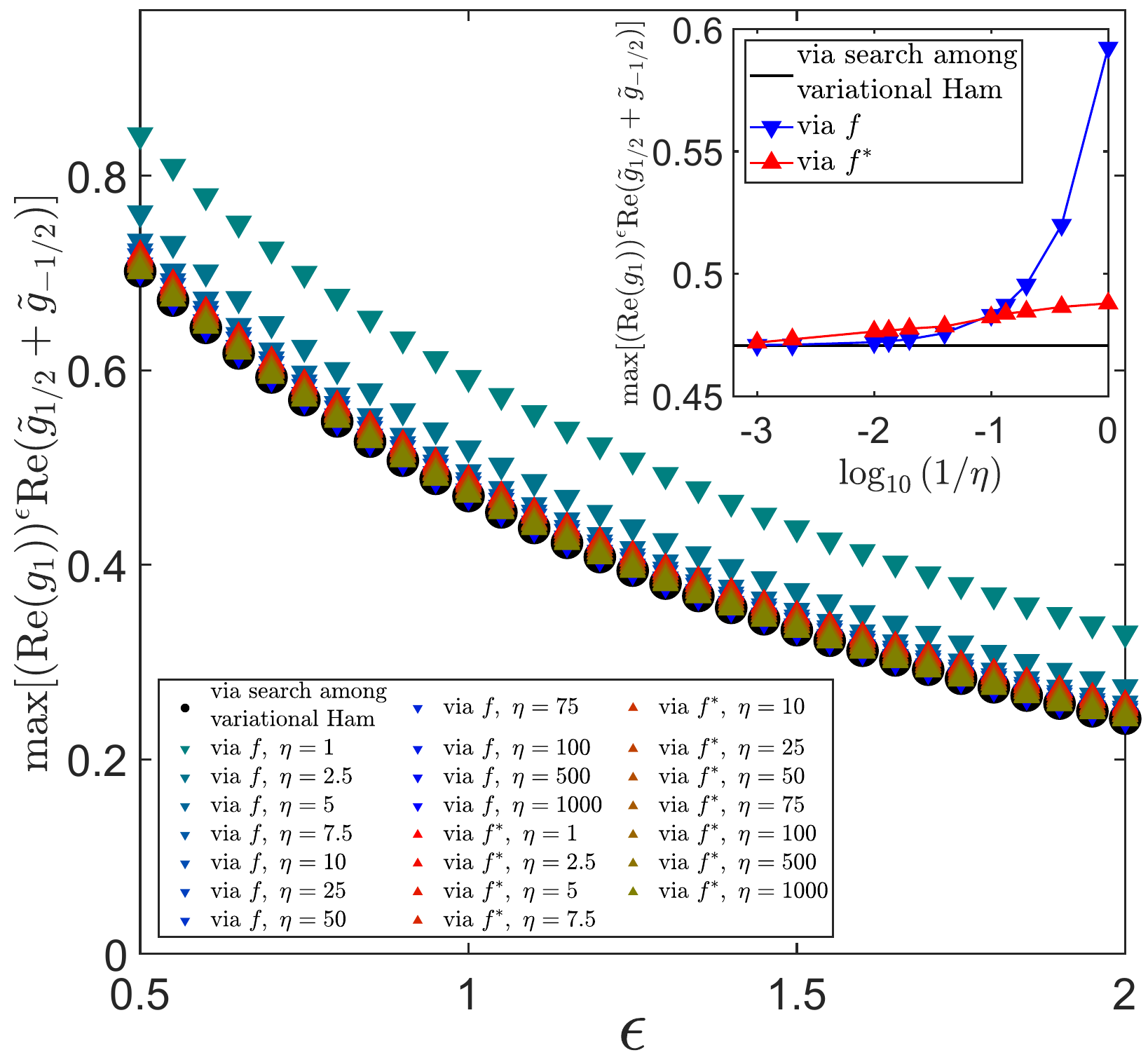}
	\caption{More details for Fig. 3 in the main text with different weight $\eta$ of the quantum constraints $f$ or $f^{*}$ are generally consistent with each other. Similar to the main text, we start from $\epsilon=0.5$ and keep track of ${g_r, \tilde g_{r'}}$ while gradually increasing $\epsilon$ for all trails. Inset: an extrapolation of the target quantity at $\epsilon = 1$ with respect to the weight $\eta$ performs well with both theory expression $f$ and ANN $f^{*}$.} \label{fig:od2}
\end{figure}

Also, we summarize in Fig. \ref{fig:od2} our results on designing quantum many-body systems to maximize $(\mbox{Re}(g_1))^\epsilon \mbox{Re}(\tilde g_{1/2} + \tilde g_{-1/2})$ in addition to Fig. 3 in the main text. As we expect, larger (smaller) $\eta$ enforces stricter (looser) quantum constraints and tends to overestimate the optimal outcomes less (more). As the quantum constraints are essentially rigid $\eta \rightarrow \infty$ for truly physical states, we may scale the target properties with respect to $\eta^{-1}$ for extrapolation, e.g., Fig. \ref{fig:od2} inset. One caveat, however, is that larger $\eta$ also reduces the manifold of expectation values we deem `physical,' thus the overall acceptance rate, leading to possible underestimations due to incomplete searches within the time and step limits, as is the case for ANN quantum constraints $f^{*}$ at large $\eta$ in Fig. \ref{fig:od2}. Therefore, throughout and careful studies are necessary to avoid misleading extrapolations and interpretations.

\section{Machine learning quantum constraints for 1D interacting spin-1/2 chains}

\subsection{Additional details on machine learning quantum constraints for 1D spin-1/2 chains}

We use the infinite matrix product state representation to represent and sample our quantum states for 1D spin-1/2 systems in the thermodynamic limit:
\begin{eqnarray}
|\Psi\rangle &=& \sum_{\cdots,s_{n-1},s_n,s_{n+1},\cdots} \cdots M^{[n-1]}_{s_{n-1}} M^{[n]}_{s_n} M^{[n+1]}_{s_{n+1}} \cdots \nonumber\\
& & \times |\cdots,s_{n-1},s_n,s_{n+1},\dots\rangle,
\label{eq:iMPS}
\end{eqnarray}
where $s_{n} = \uparrow,\downarrow$ is the spin on the $n^{th}$ site. For a translation symmetric state, we can further simplify the expression by identify $M^{[n]}_{\uparrow,\downarrow}\rightarrow M_{\uparrow,\downarrow}$, which are matrices with dimension $\chi$. The expectation values of operators defined on a subsystem of size $l_{max}$ can be evaluated straightforwardly by contract all configurations to the left and to the right of the subsystem. For our demonstrations, we set $\chi=8$ and $l_{max}=6$, which is sufficient for most scenarios away from criticality, and sample the $M_{\uparrow,\downarrow}$ matrices randomly for quantum states that we evaluate expectation values for our sample dataset.

\begin{figure}
		\includegraphics[width=1.0\linewidth]{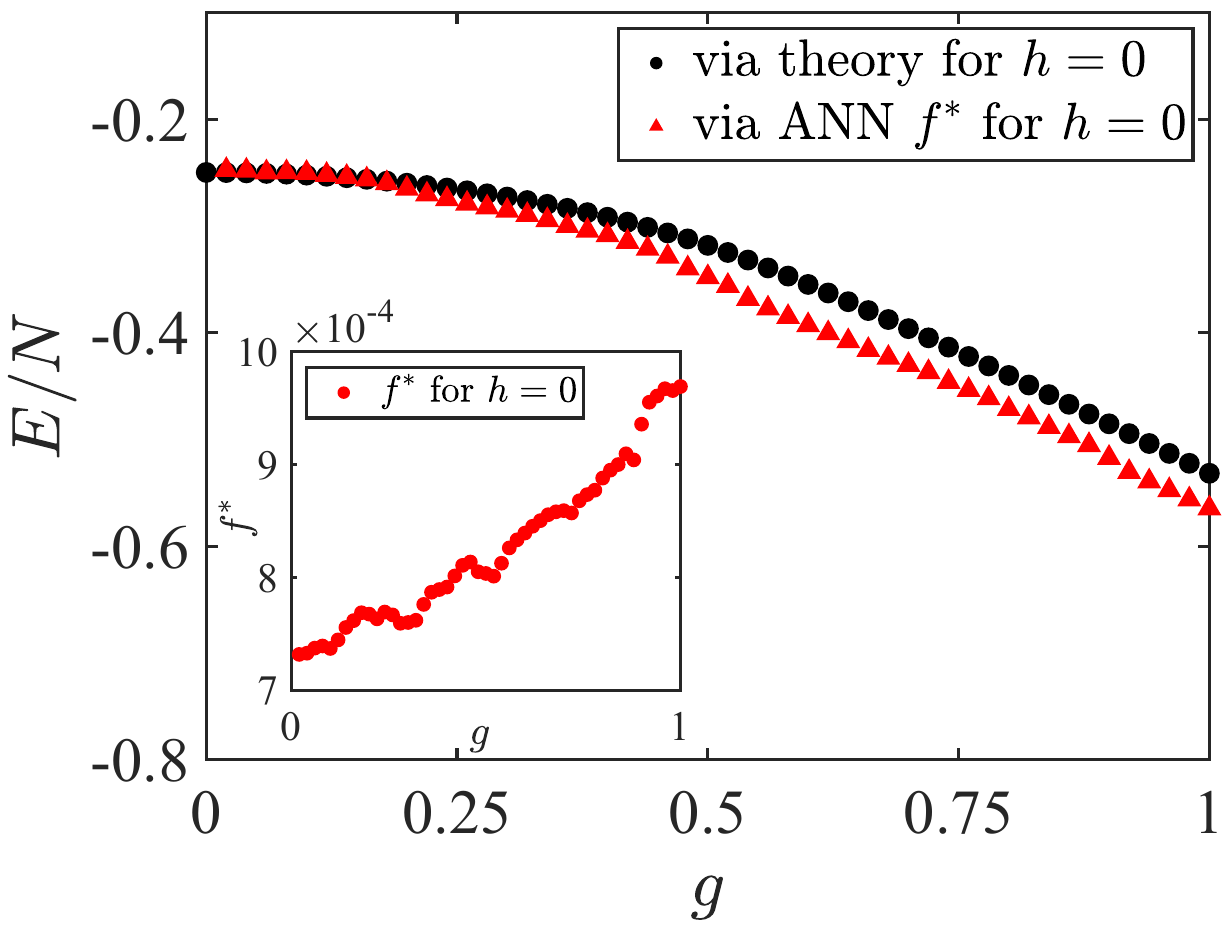}
		\caption{Similar to Fig. 4 in the main text but with different penalty weight $\eta$, the ground-state energy of the transverse field Ising model ($h=0$) in Eq. 11 obtained by the exact theory solutions \cite{LIEB1961} and the constrained minimization using ANN quantum constraints show satisfactory consistency. We set $\eta=500/3$. The inset shows the corresponding ANN outputs $f^{*}(\langle {\bf \hat O}\rangle)$ on the conditions of the quantum constraints.}\label{fig:GSE_TFI_ADAGRAD}
\end{figure}

The subsequent supervised machine learning and constrained optimization processes are similar to those of the 1D fermion insulator case, and detailed in previous sections, except that we no longer have a theoretical counterpart $f$ for benchmarks. Despite the strongly-correlated physics present in such 1D spin chains and the inclusion of higher-order operators beyond the Wick's Theorem, the data size and computational cost for obtaining $f^{*}$ and the subsequent constrained optimization are actually smaller than those of the 1D fermion insulators under current settings, since we introduce a shorter truncation length $l_{max} \ll \Lambda$ that reduces the enlisted operators. Also, importantly, the quantum states of the 1D spin chains are assumed to remain fully translation symmetric, while the quantum states in 1D fermion insulators have a bipartite unit cell that essentially doubles the active degrees of freedom. Consequently, it suffices for us to employ a similar ANN architecture and even fewer training samples for the quantum constraints $f^{*}$ on 1D spin chains.

With the quantum constraints $f^{*}$, we can evaluate the ground-state properties of 1D spin Hamiltonian in Eq. 11 in the main text via constrained optimization. Similar to Fig. 4 in the main text, we show in Fig. \ref{fig:GSE_TFI_ADAGRAD} additional data on the ground-state energy with different weights $\eta$ for the penalty. 

\begin{figure}
		\includegraphics[width=1.0\linewidth]{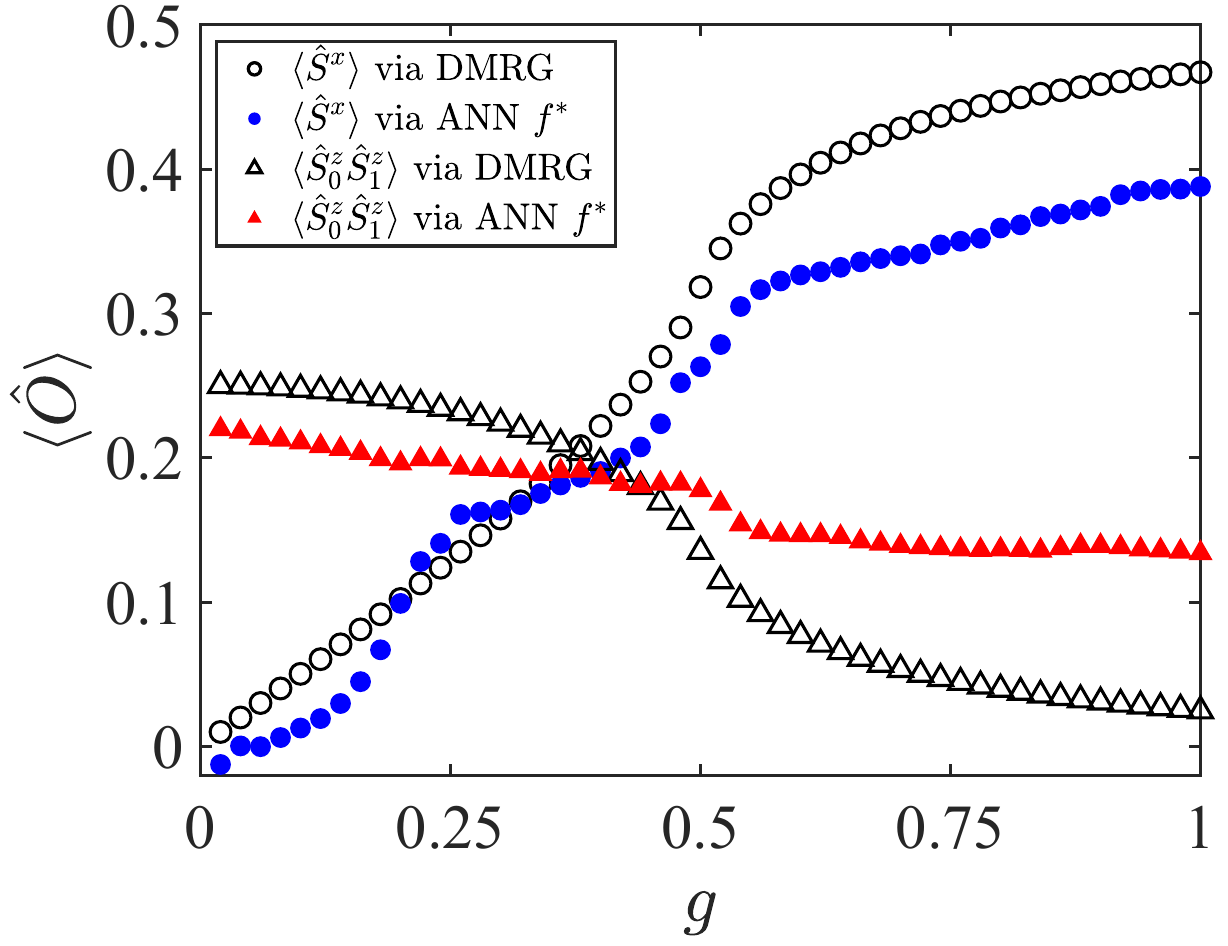}
		\caption{The expectation values of selected operators of the transverse field Ising model ($h=0$) obtained with our quantum-constraints strategy ($\eta=500/3$) as well as DMRG for benchmark.}
		\label{fig:GSE_TFIOP_ADAM}
\end{figure}

On the other hand, the measurements of expectation values of certain degrees of freedom are once again relatively more difficult for possibly similar reasons discussed in the previous section, see Fig. \ref{fig:GSE_TFIOP_ADAM}. Correspondingly, we indeed observe flatter penalty responses versus the departure in the expectation values corresponding to such operators, especially $S^x_r$, see Fig. \ref{fig:restore}. Still, our results on their expectation values follow the correct qualitative tendencies and even display some faint anomalous signatures at $g\sim 0.5$, the critical point of the transverse field Ising model ($h=0$).

\subsection{Machine learning quantum constraints for 1D spin-1/2 chains with projected wave-function samples and Variational Monte Carlo method}

In the main text, we establish quantum constraints based on Hartree Fock state and matrix product state samples, which can also generalize to and crossover various other quantum many-body ansatzes straightforwardly. In addition to the benefit of a more extensive training set, such generalizations also help reduce the bias associated with a particular ansatz and offer a more comprehensive sampling of (the relevant parts of) the Hilbert space. 

Here, we consider the following projected wave-functions for 1D spin chains:
\begin{equation}
|\Psi \rangle = \hat{P} |\Psi_{\uparrow} \rangle |\Psi_{\downarrow} \rangle,
\end{equation}
where $|\Psi_{s} \rangle$, $s=\uparrow,\downarrow$ are free-fermion states and $\hat{P}$ is a projection operator onto the one-fermion-per-site Hilbert space: $\hat{P}= \prod_i  (n_{i\uparrow} + n_{i\downarrow})(2-n_{i\uparrow}-n_{i\downarrow})$. In practice, we sample the parent free-fermion states $|\Psi_{s} \rangle$ and evaluate the expectation values $\langle \hat{\bm O} \rangle$ of $|\Psi \rangle$ through variational Monte Carlo method:
\begin{equation}
\langle \hat{O}_j \rangle = \sum_\alpha \langle \Phi | \alpha \rangle \langle \alpha|\Phi\rangle \cdot \sum_\beta \langle \alpha |\hat{O}_j| \beta\rangle \frac{ \langle \beta |\Phi \rangle}{\langle \alpha|\Phi\rangle},
\end{equation}
where $\alpha$ and $\beta$ are basis in the one-fermion-per-site Hilbert space. After generation of samples $\langle \hat{\bm O} \rangle$, we can carry out the subsequent supervised machine learning and constrained optimization as before.

\begin{figure}
		\includegraphics[width=1.0\linewidth]{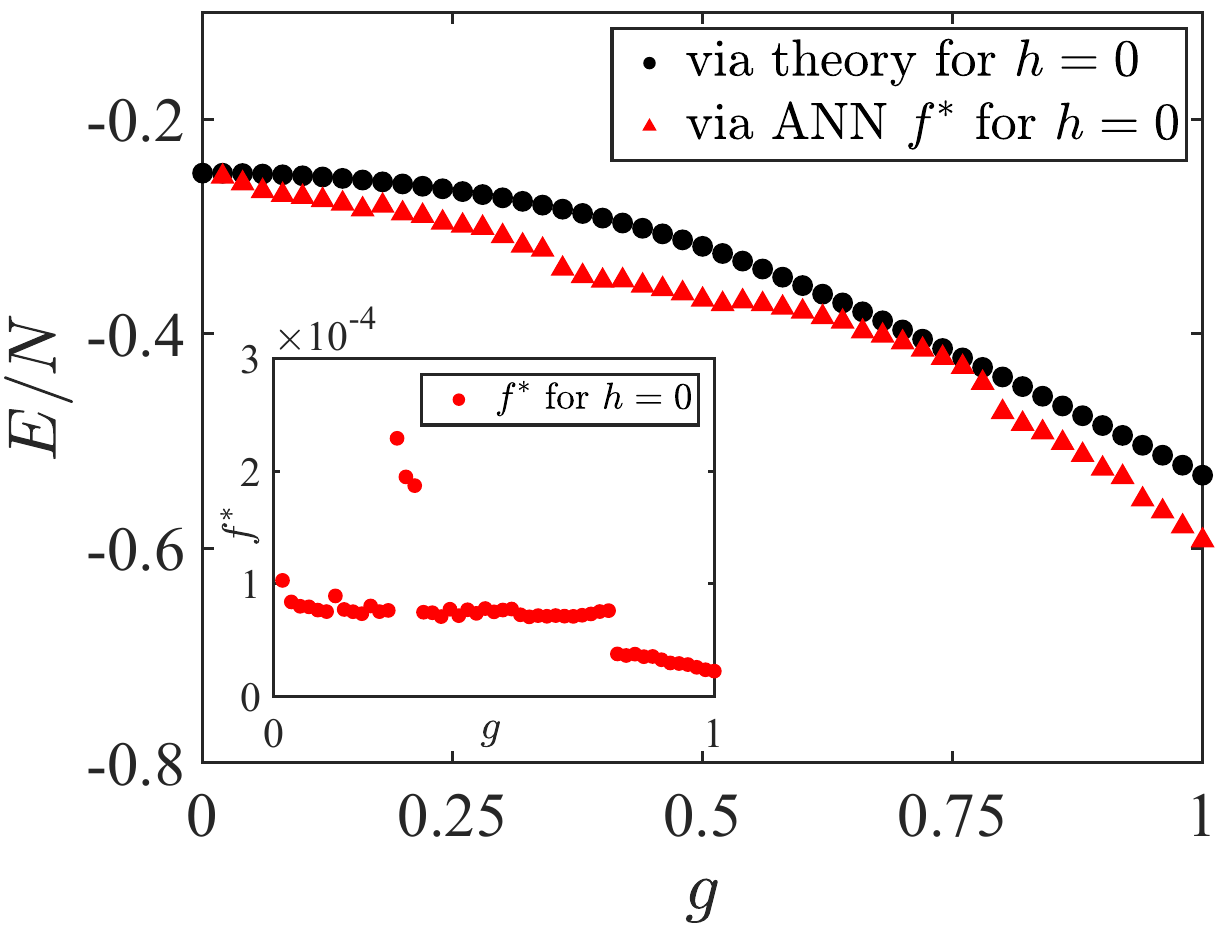}
		\caption{Similar to Fig. 4 in the main text, the ground-state energy of the transverse field Ising model ($h=0$) exhibits consistency between solutions from theory and constrained minimization with quantum constraints $f^{*}(\hat{\bm O})$. In addition to the matrix-product-state samples we consider in the main text, we train the ANNs $f^{*}(\hat{\bm O})$ on operator expectation values from projected wave-functions through variational Monte Carlo method. The corresponding system size we consider is $L=50$.}\label{fig:VMC}
\end{figure}

We summarize the results in Fig. \ref{fig:VMC}. While the overall performance is acceptable, we do not observe improvements over the performance in Fig. 4 in the main text, which only concerns samples from matrix product states. Possibly, this is because the matrix product representation is more suitable for the ground states of the transverse field Ising model; therefore, the inclusion of projected wave-function samples, though increasing the size and diversity of the training set, brings limited extra value to this particular model application. Also, the projected wave-function ansatz is real-space-based and performs on finite system sizes, e.g., $L=50$ in the current case; therefore, its incompatibility with the infinite matrix product states, though small, may confuse the ANNs upon slightly contradictory quantum constraints.

\subsection{Quantum constraints for spin-1/2 XXZ chains}

\begin{figure}
		\includegraphics[width=1.0\linewidth]{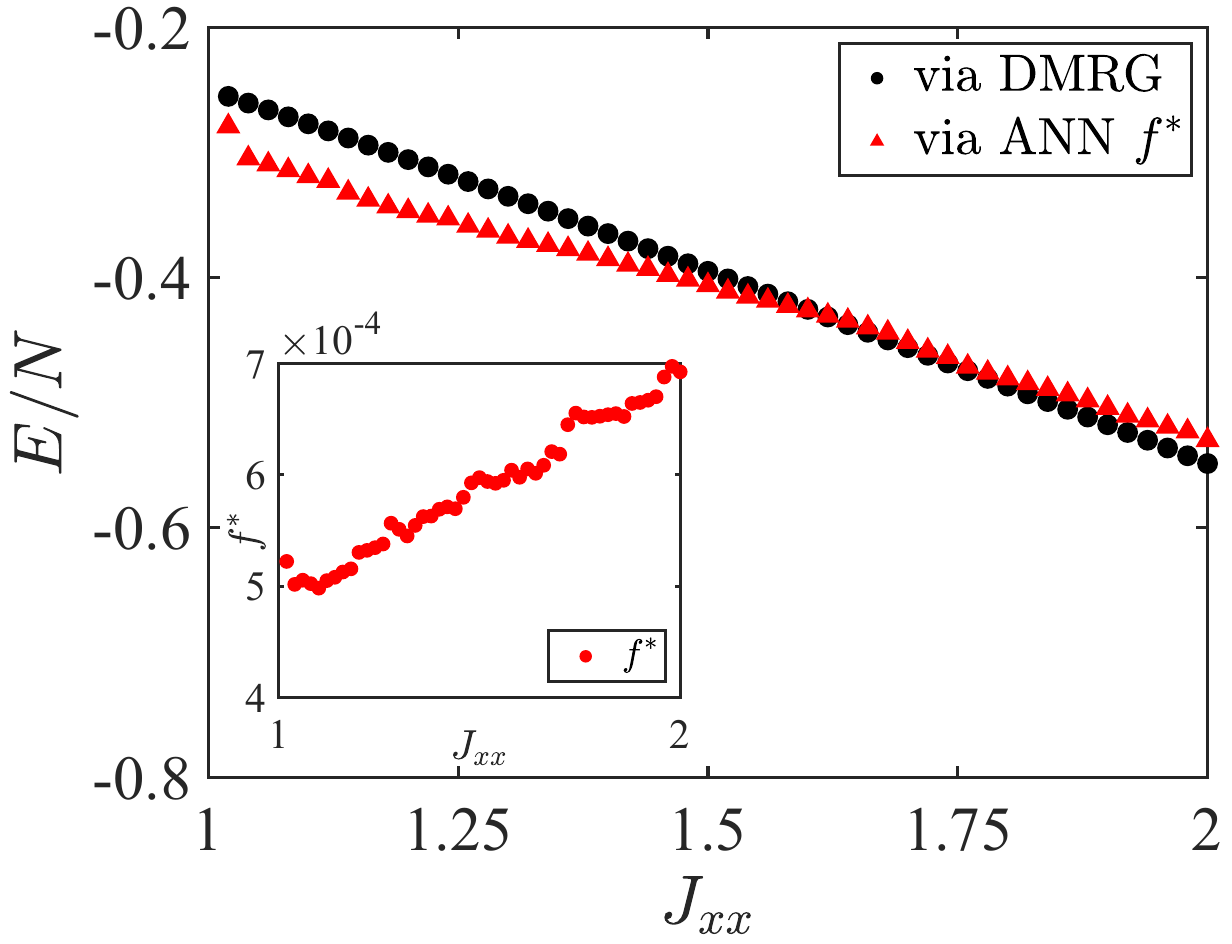}
		\caption{The ground-state energy of the spin-1/2 XXZ chain in Eq. \ref{eq:XXZ} obtained with the constrained minimization using ANN quantum constraints $f(\bf{\hat O})$ for 1D interacting spin chains in the main text exhibits satisfactory consistency with benchmark results with DMRG (the ground state should simply be the classical ferromagnetic state when $J_{xx}<J_z$). We start from the isotropic model $J_{xx}=1$ and gradually increase $J_{xx}$ for the constrained minimization with weight $\eta=1000/3$ for the penalty $f(\bf{\hat O})$. Inset: ANN $f^{*}(\langle {\bf \hat O}\rangle)$ suggests that the obtained ground state properties obey the quantum constraints.}\label{fig:XXZ}
\end{figure}

We can straightforwardly carry over our quantum constraints for 1D interacting spin-1/2 chains to analyze the ground-state properties of 1D XXZ model with ferromagnetic interactions ($J_z=1$ and $1\le J_{xx}\le2$):
\begin{equation}
    H=\sum_{i}\left[-J_{xx}(S_i^x S_{i+1}^x + S_i^y S_{i+1}^y)-J_z S_i^z S_{i+1}^z\right]. \label{eq:XXZ}
\end{equation}
In the absence of spontaneous translation symmetry breaking, the expectation value of energy per site is
\begin{equation}
E/N = -J_{xx}\langle S^x_0 S^x_1 \rangle -J_{xx}\langle S^y_0 S^y_1 \rangle -J_{z}\langle S^z_0 S^z_1 \rangle,
\end{equation}
and its ground-state value in the $N\rightarrow\infty$
thermodynamic limit is obtained via constrained optimization and summarized in Fig. \ref{fig:XXZ}.

\bibliographystyle{apsrev4-1-title}
\bibliography{refs}